\documentclass[12pt,preprint]{aastex}
\usepackage{graphicx}

\def\deg{\ifmmode^\circ\else$^\circ$\fi}
\def\arcsec{\ifmmode^{\prime\prime}\else$^{\prime\prime}$\fi}
\def\arcmin{\ifmmode^{\prime}\else$^{\prime}$\fi}
\def\fwhm{\ifmmode d\else$d$\fi}
\def\merit{capability}
\def\http{}

\newenvironment{packed_item}{
\begin{itemize}
  \setlength{\itemsep}{1pt}
  \setlength{\parsep}{0pt}
  \setlength{\baselineskip}{12pt}
  \setlength{\parskip}{0pt}
  \setlength{\lineskip}{1pt}
}{\end{itemize}}

\makeatletter
\slugcomment{}
\shorttitle{Sky Surveys}
\shortauthors{J.L.~Tonry}
\makeatother

\begin{document}

\title{An Early Warning System for Asteroid Impact}

\author{John L. Tonry$^{(1)}$}

\altaffiltext{1}{Institute for Astronomy, University of Hawaii}

\begin{abstract}
Earth is bombarded by meteors, occasionally by one large enough to
cause a significant explosion and possible loss of life.  It is not
possible to detect all hazardous asteroids, and the efforts to detect
them years before they strike are only advancing slowly.  Similarly,
ideas for mitigation of the danger from an impact by moving the
asteroid are in their infancy.  Although the odds of a deadly asteroid
strike in the next century are low, the most likely impact is by a
relatively small asteroid, and we suggest that the best mitigation
strategy in the near term is simply to move people out of the way.
With enough warning, a small asteroid impact should not cause loss of
life, and even portable property might be preserved.

We describe an ``early warning'' system that could provide a
week's notice of most sizeable asteroids or comets on track to hit the
Earth.  This may be all the mitigation needed or desired for small
asteroids, and it can be implemented immediately for relatively low
cost.

This system, dubbed ``Asteroid Terrestrial-impact Last Alert System''
(ATLAS), comprises two observatories separated by about 100~km that
simultaneously scan the visible sky twice a night.  Software
automatically registers a comparison with the unchanging sky and
identifies everything which has moved or changed.  Communications
between the observatories lock down the orbits of anything approaching
the Earth, within one night if its arrival is less than a week.  The
sensitivity of the system permits detection of 140~m asteroids (100
Mton impact energy) three weeks before impact, and 50~m asteroids a
week before arrival.  An ATLAS alarm, augmented by other observations,
should result in a determination of impact location and time that is
accurate to a few kilometers and a few seconds.

In addition to detecting and warning of approaching asteroids, ATLAS
will continuously monitor the changing universe around us: most of the
variable stars in our galaxy, many microlensing events from stellar
alignments, luminous stars and novae in nearby galaxies,
thousands of supernovae, nearly a million quasars and active galactic
nuclei, tens of millions of galaxies, and a billion stars.  With
two views per day ATLAS will make the variable universe as familiar
to us as the sunrise and sunset.

\end{abstract}


\section{Introduction}

In recognition of the hazard posed to the Earth by asteroid impact,
Congress has mandated that NASA undertake a Near Earth Object (NEO)
survey program to detect, catalog, and track NEOs of 140~m diameter
and larger.  The recent passage of a 7~m diameter asteroid 2009 VA in
Nov 2009 within only one Earth diameter emphasizes that this is a
real threat, and the fact that only a small fraction of such close
passages are detected reminds us that we are in fact in a continuous
storm of small asteroids passing close by.  In the previous year, the
Earth was struck by 2008 TC3 on 7 Oct 2008 in the Sudan.  (A
description and references can be found at the JPL NEO website,
neo.jpl.nasa.gov/news/2008tc3.html.)  Perhaps more disturbing, on
October 8, 2009 a $\sim$50~kton atmospheric explosion occurred over
Indonesia that is thought to have been caused by a $\sim$10~m asteroid
impacting the atmosphere.  (The JPL description is at
neo.jpl.nasa.gov/news/news165.html.)  
% The Earth was in the same point in its orbit as when it was struck
% by 2008 TC3 the previous year, so this may herald a new meteor
% stream that contains relatively large chunks.

The article by Asphaug (2009) reviews what is known about asteroid
populations and characteristics.  (We use the term ``asteroid'' to
mean any small solar system body, asteroid, comet, meteor, etc.)  The
population of asteroids is quite well known as a function of
brightness, usually characterized by the $H$ magnitude ($V$ band
magnitude the asteroid would have at 1~AU distance from both the Sun
and observer, viewed at opposition).  The number of main belt
asteroids goes approximately as the $-2.5$ power of size, but the
number of small ($<$200~m) NEOs reported by Brown et al. (2002) goes
more like the $-4$ power of size.  Since the arrival impact energy
goes as the cube of the asteroid size the net arrival energy is more
or less uniform per logarithmic size interval.

Conversion from observed population to surface destruction involves an
estimate of albedo to derive size (usually taken as a weighted average
of $\sim$0.14 combining $\sim$0.20 for the S-type asteroids that
predominate among NEOs and $\sim$0.05 for the C-type that are the most
numerous in the Solar System), an estimate of density to derive mass
(usually taken as $\sim$2--3~g~cm$^{-3}$ for S-type, although
ice-dominated comets have a density less than water, C-type are
$\sim$1.5~g~cm$^{-3}$, and M-type may have a density in excess of
6~g~cm$^{-3}$), an estimate of arrival velocity (typically
$\sim$15~km~s$^{-1}$, but there is a broad distribution), and an
estimate of the fraction of energy that couples through the atmosphere
to ground destruction.  An asteroid of $H$ magnitude of 22 is
therefore taken to have a diameter of 140~m and to carry about
100~Mton of kinetic energy.  Morbidelli et al. (2002) perform this
calculation in much more detail and fidelity.

The atmosphere has a surface density equivalent to about 10~m of water
so we can expect that an impactor must be {\it considerably} larger
than $\sim$10~m before a substantial fraction of its kinetic energy
reaches the ground instead of being dissipated in the atmosphere.  For
example, Melosh and Collins (2005) calculate that the $\sim$30~m iron
impactor that created the 1.2~km Meteor Crater in Arizona delivered
only $\sim$2.5~Mton to the ground of $\sim$9~Mton of arrival kinetic
energy.

Boslough and Crawford (2008) performed a detailed hydrocode
calculation of low-altitude airbursts from asteroid impact, using the
Tunguska explosion that flattened trees over an area of
$\sim$1000~km$^2$ in 1908 as a calibrator, and found that it is not a
good assumption to compare an asteroid impact to a stationary point
explosion (e.g. nuclear bomb test) at the altitude where the asteroid
explodes.  One difference is that the wake from the incoming object
creates a channel by which the explosion is directed upward, and even
a few Mton explosion will rise hundreds of kilometers into space.
Another difference is that the incoming momentum carries the fireball
and shock wave much lower than an equivalent point explosion, causing
commensurately more damage on the ground.  They also performed a
calculation of the effects of a 100~Mton stony impactor and found that
although the asteroid explodes before it hits the ground, the fireball
touches down over a diameter of 10~km with temperatures in excess of
5000~K for 10 seconds.

The 2007 impact in Carancas, Peru of a relatively small ($\sim$3~m,
1~ton kinetic energy) chondrite left a 13~m crater.  This seems to
indicate that there are mechanisms by which a small impactor can
couple significant energy to the ground, although most, like 2008 TC3
or the explosion over Indonesia in 2009, will explode harmlessly, high
in the atmosphere.

We are therefore left with some uncertainty about the frequency of
damage from asteroid impact.  The calibration by Brown et al. (2002)
of small NEOs is based on the rate of large fireballs from atmospheric
impacts and a conversion from optical to explosion energy, and this is
joined onto estimates from counts of asteroids as a function of $H$
magnitude.  The rate of impacts by large asteroids (140~m and larger)
is estimated to only one per 20,000 years or more, the rate of impacts
by 50~m, Tunguska-sized objects (5~Mton arrival kinetic energy) is
about one per 1,000 years, and the rate of 10~m (40~kton arrival
kinetic energy) impacts is about one per decade (NRC report, 2010).
These rates are probably uncertain to a factor of at least two, and
the work of Boslough and Crawford illustrates the difficulty in
predicting surface damage from the incident kinetic energy.

The NASA NEO Report (2007) found that a combination of planned surveys
by Pan-STARRS-4 and LSST could reach 83\% completeness for 140~m
diameter NEOs by 2026.  The total architecture cost was estimated
at about \$500M in FY06 dollars.  In order to speed up and improve the
detection probability, NASA found that an additional \$800M to \$1B
for either an additional LSST system dedicated to potentially
hazardous object (PHO) detection or a dedicated space imager could
bring the completion limit to better than 90\% by 2020.

This conclusion was affirmed in the recent report by the NRC report
``Defending Planet Earth: Near-Earth Object Surveys and Hazard
Mitigation'' (2010) on survey and mitigation strategies that NASA
might pursue to reduce the risk from hazardous objects, but they
stressed the severe tension between cost and survey completion
deadline, and suggested that 2030 may be a more attainable goal,
although still at high cost.  The NRC report also recognized that the
damage from relatively small asteroids in the 30--50~m range may be
greater than heretofore appreciated, and recommended that ``surveys
should attempt to detect as many 30--50-meter objects as possible''.

We believe that, while the final solution of finding, cataloging, and
tracking 90\% of asteroids of 140~m is very hard, the technology to
find most asteroids of 50~m or larger on their final approach is now
in hand.  Although this may not give us enough warning to mount a
mission to deflect the asteroid, it should give us enough warning to
know exactly where and when the impact will occur.  Lives can be
saved by moving out of the impact area or away from the tsunami
run-up, even though loss of property is unavoidable.

We describe how the construction and operation of a new sky survey
could continually scan the visible sky.  This facility, entitled
``Asteroid Terrestrial-impact Last Alert System'' (ATLAS) would use an
array of eight wide-field, fast telescopes equipped with large
detector arrays to scan the visible sky ($\sim$20,000~sq~deg) twice
per night.  Its sky completeness gives us a better than 50\% chance of
detecting any 50~m asteroid approaching from a random direction, and
its sensitivity provides three week's warning of 140~m objects 
and one week for 50~m asteroids.

The second section discusses the meaning of ``etendue'' generally, and
presents equations for ``survey \merit'' and signal to noise (SNR)
achievable from a survey instrument, even in the regime of
undersampled pixels.  This lays the foundation for evaluation of how
scientific goals can be met by a given survey implementation.  The
third section presents details of the ATLAS concept and describes how
it compares with other surveys, present and planned.  The fourth
section describes how ATLAS performs in its role of detecting
hazardous asteroids as well as other science topics.  We find that
ATLAS has some very interesting capabilities beyond early warning and
is quite complementary to other existing or planned surveys.  We
conclude with thoughts on how ATLAS could provide the seed for a
``World-Wide Internet Survey Telescope'' that could improve the
probability of detection and the warning time of approaching impactors.

\section{Etendue and Survey Design}

\subsection{Etendue and information}

The technical term ``etendue'' means the product in an optical system
of the solid angle and cross-sectional area occupied by the bundle of
light rays.  Liouville's theorem guarantees that this product is conserved
throughout the optical system, provided there is no absorption.  In
particular, at the entrance aperture, the product of solid angle seen
on the sky times the collection area is a measure of how large the
``grasp'' of an optical system is.  This ``etendue'' product is
therefore often adopted as a measure of the merit of proposed survey
systems, with some understanding that it has something to do with the
rate at which scientific value can be accrued.  

The information content in $N$ independent measurements at signal to
noise ratio (SNR) $S$ is expressible in $N\log_2(S)$ bits.  If our
science goal is to generate a catalog of independent quantities, this
might be an appropriate metric for the trade-off between quantity $N$
against quality $S$ to maximize information.
However, an important and common science goal is accumulation of SNR
for a measurement to which many correlated observations contribute.
In this case the science value lies in the SNR of the sum or other
combination of measurements that are presumed to be highly correlated,
and the information content goes as $\log_2(N^{1/2} S)$, or $\sum
{1\over2}\log_2(n_i S_i^2)$ if we do this for a number of classes of
inquiry, collecting $n_i$ objects in each class with SNR $S_i$.
Colloquially, the net SNR from averaging $N$ measurements improves
as $N^{1/2}$.

This sum, $\log(n_i S_i^2)$, is in fact the standard metric that is
used to evaluate the capability of a survey system.  It is not unique,
nor is it appropriate for all scientific goals (for example it does
not optimize detection rate of stellar occulations by hot Jupiters,
where the dependency of capability on $S_i$ is essentially a step
function at $S_i\sim200$), but it does describe the main scientific
purposes to which survey data are usually applied.  We will use the
term ``\merit'' henceforth to mean ``accumulation of $\log(n_i S_i^2)$
per unit time'' in order that ``etendue'' can be reserved for its
technical use.

In the Poisson limited case, where the variance is proportional to the
number of photons collected and $n_i$ is proportional to the solid
angle surveyed, $(n_i S_i^2)$ does not depend on how survey time is
apportioned between area coverage and depth -- \merit\ is basically
the number of photons collected from objects of class $i$ regardless
of which objects the photons come from.  There are two curbs on this
covariance, apart from the details of luminosity function or spatial
distribution.  The first arises when systematic error at extremely low
or high $S_i$ (e.g. ``read noise'' or ``flatfielding error'') slows
the growth of information from $n_i^{1/2}$ -- it is often not
practical to increase $n_i$ without bound by permitting $S_i$ to
become arbitrarily small, nor do we necessarily gain by
arbitrarily increasing $S_i$ on a single object.  The second limit
arises when $n_i$ becomes so large within a solid angle that objects
blur together -- their perceived fluxes are no longer independent,
which again limits the growth of information.  If we assign a
footprint solid angle $\omega$ to an object blurred by the point spread
function (PSF) and consider an object to have value only if no other
object's footprint overlaps its center, the maximum density of
isolated objects is achieved when the overall number density is
$\omega^{-1}$, at which point the density of non-overlapped objects is
lower by the natural logarithm base, $(e\omega)^{-1}$.

Apart from these considerations, and for known objects with sufficient
density on the sky or randomly positioned objects, maximizing $n_i$ is
tantamount to maximizing the solid angle that can be observed per unit
time.  Therefore we can consider the survey metric to be $\Omega
S_i^2$, where $\Omega$ is the survey solid angle, but remembering that
this is not valid when $S_i$ is low enough to be affected by
systematics or when the PSF and object footprint is large enough that
objects start to overlap.  The survey \merit\ is the rate at which
$\Omega S_i^2$ is accumulated.

\subsection{SNR and PSFs}

Recovery of an unresolved object's flux in the face of blurring and
noise is a finely honed art.  For uniform, independent Gaussian noise
the optimum SNR occurs by cross-correlating (often mis-named
convolving) the image with the PSF.  More generally, the optimum
cross-correlation kernel is just the Wiener filter, whose Fourier
transform (FT) depends on those of the PSF, $P(k)$, and the noise,
$N(k)$: $|P|^2/(|P|^2+|N|^2)$.  In the limit that an object is faint
compared to the noise the optimum kernel then devolves to the PSF
itself, but if the object's noise variance is significant or if the
background noise is correlated (e.g. by rebinning) the optimal kernel
becomes narrower in image space.

Note that this is true for undersampled images as well, where it is
understood that the kernel is the convolution of a ``physical PSF''
(meaning distribution of delivered flux prior to integration within
a pixel) with a detector pixel {\it with phase shift}, i.e. the
optimal kernel depends on the exact sub-pixel position where the
object lies.

The net SNR from a faint point source of total flux $f$ spread over a unity
integral PSF $P$, in the face of independent, Gaussian noise variance
per unit area (square arcsec, for example) $\sigma^2$, derived from
integration against a unity integral kernel $K$ is just
\begin{equation}
  \hbox{SNR} = {f\over\sigma} \; {\int K P \over [ \int K^2 ]^{1/2}}
             = {f\over\sigma} \; [ \smallint P^2 ]^{1/2},
\label{eq:snrpsf}
\end{equation}
where the right side expresses the SNR when the PSF is used as an
optimal kernel.

We integrated equation \ref{eq:snrpsf} for a variety of popular PSF
models, with results from the well-sampled regime given in
Table~\ref{tab:psf}.  Since $[\int P^2]^{1/2}$ for a given PSF must be
inversely proportional to the spatial scale in order to maintain unity
integral, the SNR will be proportional to $f/\sigma\,\fwhm$, where we
use the full width half maximum (FWHM) \fwhm\ as a convenient measure
of the PSF extent.  We see that aperture photometry delivers an SNR
that is about 12\% worse than PSF integration.

\noindent
\begin{table}[!ht]
\begin{center}
\caption{SNR for different PSFs}
\begin{tabular}{|lcccc|}
\hline
PSF & $\alpha_{PSF}$ & $\alpha_{circ}$ & $r_{circ}$ & Atm?\\
\hline
Gaussian         & 0.66 & 0.60 & 0.70 & N \\
Kolmogorov       & 0.57 & 0.51 & 0.71 & Y \\
Moffat           & 0.58 & 0.51 & 0.73 & Y \\
Waussian         & 0.52 & 0.45 & 0.76 & Y \\
Cubic Lorentzian & 0.40 & 0.28 & 0.84 & N \\
\hline
\end{tabular}
\label{tab:psf}
\end{center}
{\small\footnotesize Notes: The profiles are Gaussian, a Kolmogorov
  $\exp(-k^{5/3})$ profile, a Moffat (power of a Lorentzian) profile
  $(1+r^2)^{-\beta}$ with $\beta=-4.765$ recommended by Trujillo et
  al. (2001), a ``Waussian'' (wingy Gaussian)
  $(1+r^2+r^4/2+r^6/12)^{-1}$ introduced by Schechter et al. (1993)
  for DoPhot, and a cubic Lorenzian (i.e. Moffat function with
  $\beta=3/2$).  The second column is the proportionality factor
  $(\hbox{SNR}\,\sigma\,\fwhm/f)$ for a PSF
  kernel, the third column the factor for an optimal circular,
  top-hat kernel, the fourth column the optimal top-hat radius in
  units of \fwhm, and the fifth column indicates which PSF profiles
  are realistic approximations to atmospheric PSFs.}
\end{table}

We also performed these integrations into the extremely undersampled
regime, averaging SNR$^2$ over PSF position within a square pixel,
and found that a reasonable fit to the results for the middle three
(atmospheric) PSFs from Table~\ref{tab:psf} is
\begin{equation}
  \hbox{SNR} = {f\over\sigma} \; (3.5\fwhm^2+0.4\fwhm p+p^2)^{-1/2}
\label{eq:snrfwhm}
\end{equation}
where \fwhm\ is the {\it physical} FWHM, $p$ is the pixel size,
$\sigma^2$ is the background noise variance per unit area, and $f$
is the total flux in the faint point source.
Equation~\ref{eq:snrfwhm} is therefore a good approximation for
atmospheric PSFs, regardless of undersampling.
% Note that the factor 3.5 is what one would expect for well sampled
% imagery, $\fwhm{\sim}2p$ -- the ``effective FWHM'' of a pixel that one
% might add in quadrature is the about 70\% times the size of the pixel.

An application of equation~\ref{eq:snrfwhm} is selection of pixel size
for optimal SNR, trading off read noise against background noise
variance.  (For zero read noise the optimum is an infinitely small
pixel.)  If we add $\sigma_R^2/p^2$ to $\sigma^2$, where
$\sigma_R$ is the read noise per pixel, we can solve for the pixel
size that maximizes the SNR.  An approximate solution to the quartic
equation is given by
\begin{equation}
p_{opt}^2=2 \fwhm \, \sigma_R / \sigma.
\label{eq:popt}
\end{equation}
Since $\sigma_R/\sigma$ is the pixel size at which the read noise
equals the sky noise, the pixel that optimizes photometry SNR is
$\sqrt2$ times the geometric mean of the physical FWHM and the
size that balances read noise against sky noise (which depends on
bandpass and sky brightness).

\subsection{Survey design and performance}

A survey system's ability to capture photons from a source depends on
its aperture and obstruction, vignetting, filter bandwidth and
throughput, atmospheric throughput, detector quantum efficiency and
fill factor which we bundle into a single throughput number
$\epsilon$.  Operationally, we use the zeropoint of the AB magnitude
system, $5.48\times10^6$~photons~cm$^{-2}$~sec$^{-1}$~$\ln(\nu_2/\nu_1)^{-1}$,
to find that an AB magnitude of $m_0=25.10$ provides one photon per
m$^2$ per sec per bandpass of 0.2 in natural log of wavelength (a
typical width for astronomical filters).  We define $\epsilon$ as the
factor by which an actual system falls short of this ideal (or
conceivably exceeds it by using a broader bandpass), i.e. the signal
from a source of magnitude $m$ captured by an aperture of area $A$ is
\begin{equation}
f = A\; \epsilon \; t_{exp} \;10^{-0.4(m-m_0)}.
\label{eq:signal}
\end{equation}

We define the net fraction of shutter open time, including losses for
weather, daytime, instrumental failures, etc.  as ``duty cycle'',
$\delta$.  A survey system's temporal efficiency depends on the net
exposure time devoted to a given field, $t_{exp}$, adding together
however many successive dithers are deemed necessary, and the matching
overhead time $t_{OH}$ that adds all proportionate times such as read
out, slew, focus, etc.  We term the ratio $t_{exp}/(t_{exp}+t_{OH})$
the ``temporal duty cycle'' $\delta_t$.  $\delta$ can have an
interesting relation to $\delta_t$, especially when one considers
sites in Antarctica ($\delta = \delta_t$ in the winter), in orbit, or
an ``observatory'' that consists of many units at a diversity of
geographical location, but generally speaking we have control over
$\delta$ when designing a survey but control over only $\delta_t$ when
operating a survey (by changing $t_{exp}$).

Let us define the ``PSF footprint'' $\omega$ as the solid angle that
carries background noise equal to $f$/SNR for a faint point source of
flux $f$ and SNR defined by the flux measurement algorithm in use
(e.g. those in Table~\ref{tab:psf}), so that we can calculate SNR for
a given object by comparing its total flux to the noise found in this
``PSF footprint''.  Equation~\ref{eq:snrfwhm} gives this solid angle
as $\omega = (3.5\fwhm^2+0.4\fwhm p+p^2)$ for the case of PSF-matched
photometry with a atmospheric seeing profile, independent noise, but
not necessarily well sampled.

If $\mu$ is the sky brightness per square arcsecond,
the noise variance that the signal contends with is
\begin{equation}
\sigma^2 = A\; \epsilon \; t_{exp}\; 
\left( \omega \; 10^{-0.4(\mu-m_0)} + 10^{-0.4(m-m_0)}\right) + f_{R}^2,
\label{eq:noise}
\end{equation}
where $f_R^2$ is the readout variance over $\omega$'s worth of pixels:
$f_R^2 = \sigma_R^2\;p^{-2}\;\omega$ for a read noise of $\sigma_R$
e$^{-}$ and pixel size $p$ arcsec.  \footnote{Note that the term involving the
object's magnitude $m$ itself is somewhat notional -- not only does the
weighting involve $\int P^3$ for the case of a PSF kernel, but
the optimal kernel would change for objects that are brighter than noise.
Its presence serves to remind us that SNR depends on the Poisson
statistics of the flux from the object itself.}

Equations \ref{eq:signal} and \ref{eq:noise} provide the square of the
signal to noise ratio, $S_1^2$ for at this particular magnitude:
\begin{equation}
S_1^2 = A\; \epsilon \; t_{exp} \; \omega^{-1}\; 
10^{+0.4(\mu-m_0)} \;
10^{-0.8(m-m_0)} \;
\left[1 + 10^{-0.4(m-m_{sky})} + f_R^2\;f_{sky}^{-1}\right]^{-1}, 
\end{equation}
where $m_{sky}$ is sky magnitude within $\omega$,
$m_{sky}=\mu-2.5\log\omega$,
and $f_{sky}$ is equivalent flux in e$^-$,
$f_{sky}=A\epsilon t_{exp} 10^{-0.4(m_{sky}-m_0)}$.
In this equation and below, the
term in square brackets is approximately unity when the sky noise
dominates the object's photon statistics and the read noise; we
include it here for completeness, but drop it henceforth for clarity.
It can be reintroduced if the read noise or object photon noise is
significant with respect to the background noise, and it causes the
turn-down in the curves of Figure~\ref{fig:surveys}

The \merit\ metric defined above includes a factor for the surveyed
solid angle.  The cadence time $t_{cad}$ to carry out $t_{exp}$ worth
of integration over a survey solid angle $\Omega$ is related to the
field of view solid angle $\Omega_0$ and duty cycle by
\begin{equation}
\Omega \; t_{cad}^{-1} = \Omega_0 \; t_{exp}^{-1} \; \delta.
\end{equation}
Therefore the \merit\ function at magnitude $m$ is
\begin{equation}
{S_1^2\;\Omega\over t_{cad}} = {A\; \Omega_0 \; \epsilon \; \delta
\over \omega} \; 10^{+0.4(\mu-m_0)} \;10^{-0.8(m-m_0)},
% \left[1 + 10^{-0.4(m-m_{sky})} + {f_R^2\over f_{sky}}\right]^{-1}, 
\end{equation}
This includes the $A\Omega_0$ term commonly called ``etendue'', but
also the dependence on $\omega$, $\epsilon$, $\delta$, and $\mu$
that are crucial to the real SNR gathering capability of a system.
Rewriting the system-fixed parameters as an overall system \merit\ $M$,
equation~\ref{eq:etendue} reveals how the survey choices of cadence,
SNR, survey area, and magnitude can be traded off against one another.
\begin{equation}
M = {A\; \Omega_0 \; \epsilon \; \delta \over \omega} \; 10^{+0.4(\mu+m_0)} = 
{ S_1^2 \; \Omega \; \over t_{cad}} \;10^{+0.8m}.
% \left[1 + 10^{-0.4(m-m_{sky})} + {f_R^2\over f_{sky}}\right].
\label{eq:etendue}
\end{equation}

\begin{figure}[!hb]\begin{center}
\centerline{\includegraphics[scale=0.5]{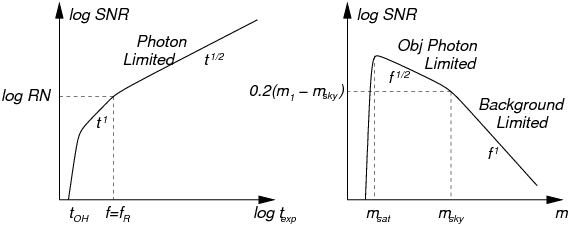}}
\caption{SNR, illustrated by cuts at constant magnitude and $t_{exp}$, falls
  precipitously when $t_{exp}$ approaches the overhead time $t_{OH}$
  ($\delta_t\ll1$) or
  the magnitude approaches the saturation limit $m_{sat}$ (which depends on
  $t_{exp}$ of course), falls quickly with exposure time when the flux is
  less than the read noise $f_R$, and transitions between photon
  and background limited when the magnitude becomes fainter than the
  background magnitude $m_{sky}$ within a PSF.}
\label{fig:observability}
\end{center}\end{figure}

Taking a logarithm, survey-variable parameters on the right
add to the (nearly) constant left hand side:
\begin{equation}
\log M = 0.8m - \log t_{cad} + \log \Omega + 2 \log S_1.
% + \log[1 + 10^{-0.4(m-m_{sky})} + f_R^2/f_{sky}].
\end{equation}
This relation forms a surface in the observability space of magnitude
$m$, SNR, solid angle, and cadence interval that is accessible for a
particular survey \merit.  The left hand side is not strictly fixed;
changing $t_{exp}$ affects $\delta_t$ and $\delta$ as well as the
(dropped) term in square brackets if the read noise is not negligible,
and the term in square brackets also contributes if the object is
brighter than the background.  A sketch of how the left hand side is
affected is illustrated in Figure~\ref{fig:observability}.  Apart from
this, the ``observability surface'' is a plane in log space.

As argued in the section above, most science value is not
changed by tradeoffs that keep the product $\Omega S_1^2$ constant.
In practice scientists tend to set $S_1$ at a fixed, minimum value for
which systematics are not compromising the SNR and then maximize
survey solid angle $\Omega$.  For moving object detection $S_1$ might
be 5; for Type Ia supernova light curves $S_1$ might be 30 at peak;
for planetary occultations $S_1$ might be 200.  This ``science value
level set'' or SNR operating point provides second constraint in
observability space for a survey.  Therefore there are really only two
independent parameters for setting a given survey's operation for a
given \merit: for a given magnitude the cadence time dictates the
solid angle.

The density of various types of objects and application of science
values can now optimize the overall survey \merit.  For example
if value lies in detection of orphan afterglows of gamma ray bursts
(GRB) we may
choose to spend our $M$ \merit\ in short $t_{cad}$ and large $\Omega$ at
the expense of $m$.  If value lies in detection of planetary
occultations of stars we cannot give up $S_1$ or $t_{cad}$ and
therefore may make compromises in $m$ or $\Omega$.  
% Of course the survey equipment also must be designed to support 
% short $t_{cad}$, for example by making $t_{OH}$ as small as possible.
Searching for
solar system objects would emphasize $m$ and $\Omega$ at the cost of
minimal $S_1$ and allowing $t_{cad}$ to grow to the linking confusion
limit.  General purpose surveys such as PTF, Pan-STARRS, and SkyMapper
strive to maximize \merit\ $M$ generally, but then dedicate portions of time
(subdivide $\delta$) to different locations in the observability
surface according to different science goals.  LSST has claimed to be
able to maximize science value at a fixed location on the
observability surface, but of course it is straightforward to move on
the surface or split time into different surveys should that prove
desirable.

\section{ATLAS}

Spaceguard has discovered most NEOs larger than 1~km, and has
determined that none will strike the Earth in the foreseeable future.
The NRC report (2010) estimates that the remaining fatality rate is
bimodal as a function of impactor size, with a $10^{-6}$~yr$^{-1}$
probability of impact by a 1--2~km object that would cause 50 million
deaths (averaging over possible impact locations), and a
$10^{-3}$~yr$^{-1}$ probability of impact by a $\sim$50~m object that
would cause an average of 30,000 deaths.  The $H$ magnitude of a 50~m
asteroid is 24 or fainter, and for a typical phase function the actual
magnitude at 1~AU distance will be $>$25--26.  This suggested to us
that surveying at a much smaller distance than 1~AU would make sense,
and by definition any Earth-impacting asteroid will be present shortly
before impact at a small distance.  Choosing one week as a minimum
warning interval for civil defense against a limited explosion and
three weeks warning as necessary for a city-devastating explosion, we
were surprised to discover that this places rather modest requirements
on limiting magnitude, although it does require isotropic vigilance.

It is possible to achieve the requisite sensitivity over half the sky
with survey hardware that is more or less off-the-shelf and of modest
cost.  We have proposed ATLAS to NASA for construction and two years
of operation; fundamentally the system is equivalent
to a telescope of 0.5~m aperture with a 40~sq~deg field of view,
subjected to an effective PSF of 3.8\arcsec, with bandpasses twice
as wide as an SDSS filter.  By comparison, the Palomar Transient
Factory uses a 1.2~m aperture, an 8~sq~deg field of view, SDSS
filters, and enjoys $\sim$2.2\arcsec\ seeing for a very comparable
\merit.

The NASA proposal implements ATLAS using eight Takahashi astrographs
of 0.25~m aperture and 0.7~m focal length that each provide a
20~sq~deg field of view.  These telescopes are small enough that there
are a number of equatorial mounts available commercially that can
carry more than one telescope.  We believe that a fully equipped
telescope with focusser, filters, shutter, camera adapter and mount
should cost about \$50k.

The cameras for ATLAS each have a 4$\times$4~k pixel focal plane,
taking advantage of an existing inventory of 2$\times$4~k CCDs with
15~um pixels.  Although the pixel size is not optimal in the sense
above (a 10~um pixel provides about 0.1 mag more sensitivity in
moderate seeing), a pair of those CCDs could be mounted in a cryostat
and equipped with a controller for a unit cost that we again believe
will be about \$50k (since there is no detector cost).

ATLAS consists of a set of eight of these telescope and camera units,
and reaches an interesting survey \merit\ level, while remaining
cost effective.
While subject to further optimization, the design reference calls for
\begin{packed_item}
\item Exposures of $t_{exp}=30$~sec with $t_{OH}=5$~sec and 10~e$^-$ read
  noise, using broad filters that are approximately $g$+$r$ and $r$+$i$.
  (The science program of finding asteroids calls for the broadest
  possible filters; the other science programs benefit from color
  information.) 
\medskip
\item Four unit telescopes are clustered on a common mount within an
  observatory, and the other four in an identical observatory
  separated by $\sim$100~km in order to obtain good parallaxes to
  0.1~AU, enabling instantaneous alerts for approaching objects as
  well as providing the crucial function of weeding out false alerts
  from space junk.  The ``blue'' observatory uses the $g$+$r$ filter
  set; the ``red'' observatory the $r$+$i$ filters.  At each
  observatory telescopes are used as two co-aligned pairs, thereby
  providing 40~sq~deg of instantaneous field of view at twice the
  aperture of a single telescope and twice the throughput of a single
  SDSS filter.  The two observatories synchronize their pointing and
  observations exactly.
\medskip
\item Both observatories cover the entire, visible sky (20,000~sq~deg)
  twice per night, visiting each point with a time separation of about
  1 hour (RA permitting) in order to obtain unambiguous tracklets of
  moving objects.
\end{packed_item}

Although the 15~um pixels subtend 4.4\arcsec\ and are therefore
considerably undersampling the PSF, a detailed calculation of the
expected sensitivity is promising.  A single, moonless exposure in
either bandpass by each of the telescopes reaches SNR 5 at $V=19.1$
for a solar spectrum.  The seeing assumed for this was 1.5\arcsec, but
the SNR is relatively insensitive because of the undersampling and the
contribution from optics blur.  The sum of the images from a
co-aligned pair of units therefore yields SNR 5 at $V=19.5$.  The
combination of the observations from the two co-aligned pairs at the
two observatories provides SNR 5 at $V=19.9$.  (The two filters are
chosen to provide the same SNR for a solar spectrum.)

ATLAS's performance on objects fainter than this depends on the
details of the object and how the observatories can communicate.  Our
design calls for each observatory to have a cluster of computers that
can align and co-add images, and can reliably detect objects at SNR
3.7 (with false alarms).  We demand that the observatories have at
least enough bandwidth that they can share detections, so as to
confirm SNR 3.7 detections and reach SNR 5 at $V=19.9$.

We cannot expect to detect moving objects (or objects closer than 0.05
AU with a significant parallax) much fainter than $V\sim20$ since they
will not align on successive images.  However, stationary objects that
are observed twice per night will be detected at SNR 5 at $V\sim20$ in
both the red and blue bandpasses, or SNR 5 at $m=20.35$ in a
combination of red and blue images.  Obviously detections of
stationary objects can continue to fainter magnitudes by stacking many
night's observations until systematics dominate.

With no defocus, the cameras will saturate at $V\sim12.5$ (blue) and
$V\sim13$ (red).  We intend to equip each observatory with a pair
of high-end digital SLR cameras to provide 5 color photometry to a
limiting magnitude of $V\sim6$, so as to be able to monitor brighter
stars and extend the dynamic range for very bright transients.

\section{Ongoing and Planned Surveys}

There are many past, ongoing, and future sky surveys for a variety of
purposes.  Table~\ref{tab:surveys} shows the basic design choices made
by a set of successful efforts and some proposed ones.

The most productive asteroid and near Earth object (NEO) search
programs are currently the Catalina Sky Survey (Larson et al., 2003;
\http\ www.lpl.arizona.edu/css), LINEAR (Stokes et al., 2000;
\http\ www.ll.mit.edu/LINEAR), and Spacewatch (McMillan et al., 2006,
Larson et al., 2007;
\http\ spacewatch.lpl.arizona.edu).  The JPL website (neo.jpl.nasa.gov)
attributes 73\% of asteroid discoveries in 2009 to Catalina (60\% of
NEOs), 14\% to LINEAR (28\% of NEOs), and 8\% to Spacewatch (4\% of
NEOs).  (Spacewatch is now spending a greater fraction of time on
follow-up rather than discovery.)

``Pi of the Sky'' (Malek et al, 2009; \http\ grb.fuw.edu.pl) is a
representative GRB search program.  RAPTOR (Vestrand et al., 2003;
\http\ www.raptor.lanl.gov) is another interesting
example of GRB and other transient search, but is not listed in
Table~\ref{tab:surveys}.  These projects put a high premium on
rapid cadence and rapid follow up capability, at the cost of limiting
magnitude. 

``SuperWasp'' (Pollacco et al. 2006; http: //www.superwasp.org) and
``HAT-South'' (Bakos et al. 2009;
\http\ www.cfa.harvard.edu/$\sim$gbakos/HS) are examples of surveys
searching for planetary occultations.  Such surveys must work at very
high SNR at fast cadence, again at the expense of limiting magnitude,
but their science does not lack for stars of suitable brightness.
HAT-South is particularly interesting for comparison with ATLAS
because there are marked similarities in the equipment, but the
science for HAT-South and ATLAS lives in different locations in
observability surface.

The Palomar Transient Factory (PTF) (Law et al., 2009;
\http\ www.astro.caltech.edu/ptf) has dedicated time to different
search strategies for optical transients such as supernovae.  We list
the properties of the ``5 day'' portion of their survey.

Pan-STARRS1 (Burgett \& Kaiser, 2009; \http\ pan-starrs.ifa.hawaii.edu) and
SkyMapper (Keller et al., 2007; \http\ rsaa.anu.edu.au/skymapper)
seek to perform surveys of the entire northern and southern skies to
unprecedented depths.  Both surveys are optimized to find
transients and moving objects.  A portion of Pan-STARRS1 is dedicated
to a ``3pi'' survey of 3/4 of the sky, revisiting once every 3 months;
another portion to a ``Medium Deep'' (MD) survey of about 40~sq~deg
revisited each day to a substantial depth.  We
list both Pan-STARRS surveys in order to illustrate how more or less
equal resources (\merit) can be placed at rather different places on
the observability surface.  Pan-STARRS is intended to be a replicable
system, with a goal of four units (PS4) sited on Mauna Kea.

The Large Synoptic Survey Telescope (LSST) (Ivezic et al., 2008;
\http\ lsst.org) is proposed to survey the visible sky on a
few day cadence with an 8~m telescope, projected to reach at least
a magnitude fainter than Pan-STARRS1 at a much faster cadence.

\noindent
\begin{table}
\caption{Sky Survey Design and Performance}
\begin{tabular}{|lrrrrrrrrrrr|}
\hline
Program  &$A$ &$\Omega_0$&$\omega$&$m_{sky}$&$\delta_t$&$S_1$&$m$&$n_c$&$t_{cad}$&$\Omega$&$\log M$\\
\hline
\multicolumn{12}{|c|}{NEO search} \\
\cline{4-8}
Spacewatch & 0.51 &  2.9&   16&  18.0& 0.50&   3&  21.7&   1&  0.5&    150&  20.8\rule{0pt}{16pt}\\
LINEAR     & 1.2  &  2.0&   30&  17.3& 0.80&   4&  19.0&   1&  0.3&   2400&  20.4\\
Catalina   & 0.27 &  8.2&   41&  17.0& 0.50&   4&  19.5&   1&  0.4&    800&  20.1\\
\hline
\multicolumn{12}{|c|}{GRB counterparts} \\
\cline{4-8}
PioftheSky & 0.13&   484&25000&  10.0& 0.83&   5&  14.5&   1&  0.01&  6400&  18.8\rule{0pt}{16pt}\\
\hline
\multicolumn{12}{|c|}{Planet occultations} \\
\cline{4-8}
SuperWASP  & 0.15 &   61&11000&  10.9& 0.88& 100&  12.9&   1&  0.02&  3900&  19.7\rule{0pt}{16pt}\\
HAT-South  & 0.46 &   16&  128&  15.7& 0.92& 100&  14.0&   1& 0.004&   128&  20.5\\
\hline
\multicolumn{12}{|c|}{Sky Survey and Transients} \\
\cline{4-8}
PTF-5day   & 0.85 &  7.8&   16&  18.0& 0.67&   5&  21.4&   2&     5&  3200&  21.3\rule{0pt}{16pt}\\
SkyMapper  & 1.1  &  5.2&    8&  18.7& 0.88&   5&  22.4&   6&   270& 20000&  21.2\\
PS1-3pi    & 1.8  &  7.5&  3.7&  19.6& 0.75&   5&  23.3&   5&    90& 20000&  22.4\\
PS1-MD     & 1.8  &  7.5&  3.7&  19.6& 0.98&   5&  24.7&   5&     4&    45&  22.2\\
\hline
\multicolumn{12}{|c|}{Proposed Surveys} \\
\cline{4-8}
ATLAS      & 0.29 &   20&   46&  16.9& 0.86&   5&  19.9&   2&   0.7& 20000&  21.8\rule{0pt}{16pt}\\
PS4-3pi    & 7.1  &  7.5&  3.0&  19.8& 0.92&   5&  23.6&   5&    10& 20000&  23.6\\
LSST       & 35   &  9.6&  2.9&  19.8& 0.88&   5&  24.5&   2&     3& 10000&  24.5\\
\hline
\end{tabular}
{\small\footnotesize
Notes: $A$ is the net aperture in m$^2$, including obscurations and
the number of units; $\Omega_0$ is the solid angle in deg$^2$ per
exposure; $\omega$ is the noise equivalent PSF area in arcsec$^2$ as
discussed in the text; $m_{sky}$ is the magnitude collected within
$\omega$ when the sky brightness is $\mu=21$ per sq~arcsec; $\delta_t$
is the exposure duty cycle $t_{exp}/(t_{exp}+t_{OH})$; $S_1$ is the
SNR achieved at magnitude $m$ which includes the coadded sensitivity
from the contributions of $n_c$ colors; the cadence times $t_{cad}$ in
days have been adjusted for 2/3 clear weather; $\Omega$ is the actual
solid angle in deg$^2$ surveyed in the cadence time $t_{cad}$;
``Catalina'' is only the 0.7~m Schmidt, its combination with the
Siding Spring Survey and the Mt. Lemmon Survey almost double the total \merit;
``PS4-3pi'' presumes a 100\% $3\pi$ survey for PS4, but the etendue
may be split as with PS1; ``LSST'' is based on current suggestions for
a 2-color, 2$\pi$ survey which is estimated to reach $m=24.5$ in 3
days, including weather.}
\label{tab:surveys}
\end{table}

Table~\ref{tab:surveys} also shows how well these surveys perform and
their log \merit, generated from equation~\ref{eq:etendue}.  Actual
performance only correlates loosely with $A\Omega_0$: background,
efficiency, and duty cycle are very serious factors.  Therefore the
\merit\ is best calculated from the {\it right} hand side since
limiting magnitude at some estimate of SNR and the overall cadence of
covering a planned solid angle is generally well reported.  For
systems which are not yet operational we take their estimated SNR and
magnitude at face value, but delivered $\epsilon$, $\delta$, and
especially $\omega$ may fall short of pre-operational claims.

\begin{figure}[!hb]\begin{center}
\centerline{\includegraphics[scale=0.25]{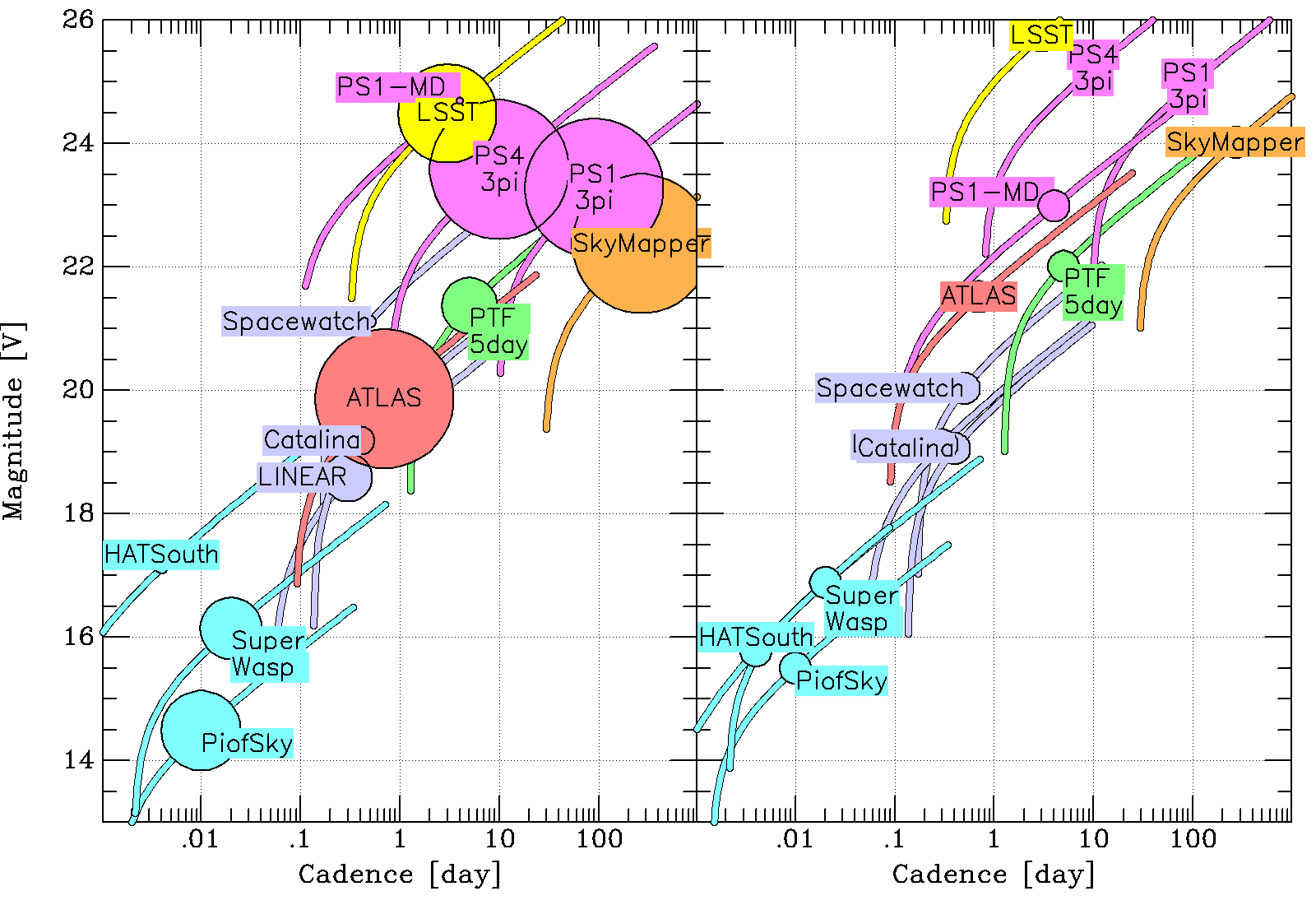}}
\caption{The various system's capabilities are shown at common SNR = 5
  in the magnitude-cadence plane on the left.  The area of the symbol is
  proportional to the solid angle $\Omega$ surveyed in that cadence
  time, and the lines illustrate the observability surface in
  $m$ and $t_{cad}$ at fixed $\Omega$ and SNR.  However,
  putting all surveys at a common SNR 5 does some injustice to the
  science they seek to achieve.  The rolloff in magnitude occurs when
  $t_{exp}$ becomes comparable to $t_{OH}$ (the square bracket term).
  On the right, the various systems capabilities are shown in the
  magnitude-cadence plane at common SNR = 5 {\it and} assuming they
  trade off sensitivity for coverage to $\Omega$ = 1000~sq~deg.}
\label{fig:surveys}
\end{center}\end{figure}

Different survey choices can trade off $m$, $S_1$, $t_{cad}$, and
$\Omega$ against one another on the observability surface;
Figure~\ref{fig:surveys} shows a cut at constant $S_1=5$ in the
$m-t_{cad}$ plane, with point area proportional to $\Omega$, and
another cut at $S_1=5$ and $\Omega=1000$~sq~deg.
The cadence time is defined as the mean time required to survey
$\Omega$, but the survey may include a great deal of valuable temporal
sampling when the survey includes multiple exposures.  For example,
the PS1-3pi survey is specifically designed to detect moving objects
with pairs of exposures on a $\sim$15~min interval.  It is therefore
not safe to conclude that $t_{cad}$ listed in Table~\ref{tab:surveys}
is the shortest time interval for detection of motion or variability.

It is instructive to examine how the ATLAS proposal differs from
HAT-South and Pan-STARRS1.  ATLAS is using a very similar approach to
the HAT-South project, even to the extent of both using four Takahashi
telescopes on common mounts, each feeding a $4\times4$k camera.  ATLAS
gains factors in \merit\ from $\epsilon$ ($\times$4), $\omega$
($\times$3), $\Omega$ ($\times$1.5), but loses in $A$ ($\times$0.7)
for a net gain of about an order of magnitude.  Pan-STARRS1 and ATLAS
have nearly the same product of $A\Omega_0\epsilon$ (collect photons
at the same rate), but of course Pan-STARRS1 has about an order of
magnitude higher \merit\ than ATLAS because $\omega$ is so much
smaller.

It is not worthwhile to try to split hairs about which survey is the
``best'' or most capable; many of the parameters in the table above
are rough enough that it is not possible to make an accurate
comparison.  Even more important, the table fails to clarify all the
factors which make the various surveys especially well tuned for the
science they are trying to do.
However, ATLAS does occupy an important portion of design space.  It
is an order of magnitude faster than existing NEO surveys, it reaches
much fainter magnitudes than the other sub-day cadence surveys, and it
is unique in surveying the entire sky several times per night.  While
some of the other systems could move at constant \merit\ to cover the
entire sky nightly, they would not be able to do so at nearly the
sensitivity of ATLAS.

ATLAS is complementary to general surveys such as Pan-STARRS,
SkyMapper, and LSST.  Like these it covers most of the sky, but
it offers a much faster cadence at the cost of less sensitivity, fewer
colors, and less resolution.  As described in the next section, there
is a great deal of science to be found in this brighter, faster regime
of discovery space in addition to the primary mission of finding
asteroid{}s approaching the Earth.

\section{Science with ATLAS}

\subsection{Asteroid Impacts}

ATLAS is first and foremost a system to warn of incoming objects
that might hit the Earth.  ATLAS is not optimized to find
objects at 1 AU and $H\sim20$, many of which are known and will
never strike the Earth.  Other systems,
notably Pan-STARRS and eventually LSST, have the leisure to find and
catalog these better.

Fortunately the interval between collisions of the Earth with an
object of $\sim$50~m or larger is many centuries, the impact at
Tunguska in 1908 notwithstanding.  However, the cumulative frequency
of Earth impacts as a function of the size of impactor has been
estimated by Brown et al. (2002) as
\begin{equation}
N(>D) = 37\,\hbox{yr}^{-1} \; \left( D \over 1\hbox{m} \right)^{-2.7} 
\end{equation}
so we can expect an
impact of a 20--30~m asteroid once per century, a nearly Mton-class
explosion.  Although most of the incident energy will be dissipated
high in the atmosphere, we have already discussed the evidence that it
could cause significant damage on the ground as well.

%%%%%%%%%%%%%%%%%%%%%%%%%%%%%%%%%%%%%%%%%%%%%%%%%%%%%%%%%%%%%%%%

\begin{figure}[!hb]\begin{center}
\centerline{\includegraphics[scale=0.25]{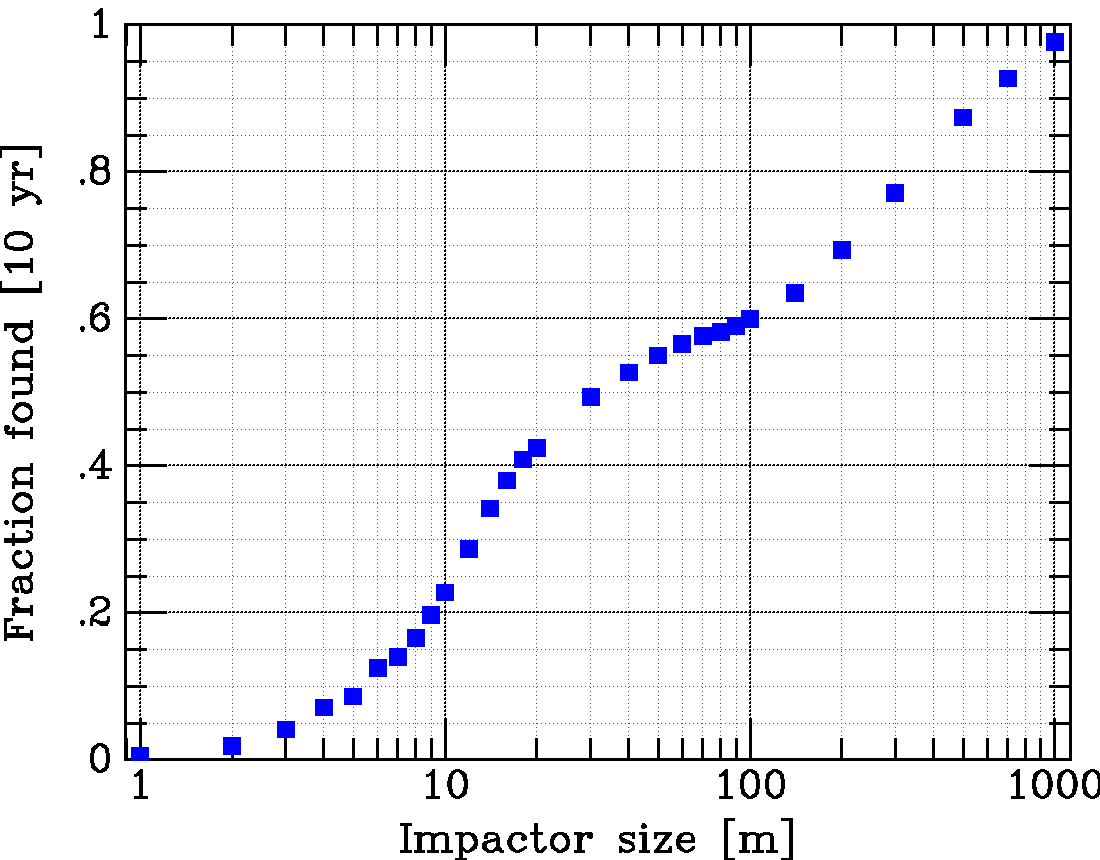}}
\caption{The fraction of impactors that ATLAS detects before
  collision is shown as a function of asteroid size for a survey of 10
  years duration.  The kink at
  $\sim$20~m occurs when a significant fraction has warning time
  greater than one day, and the kink at $\sim$140~m is caused by
  an increasing fraction of greater than one orbit warning times.
}
\label{fig:warnfrac}
\end{center}\end{figure}

\begin{figure}[!hb]\begin{center}
\centerline{\includegraphics[scale=0.18]{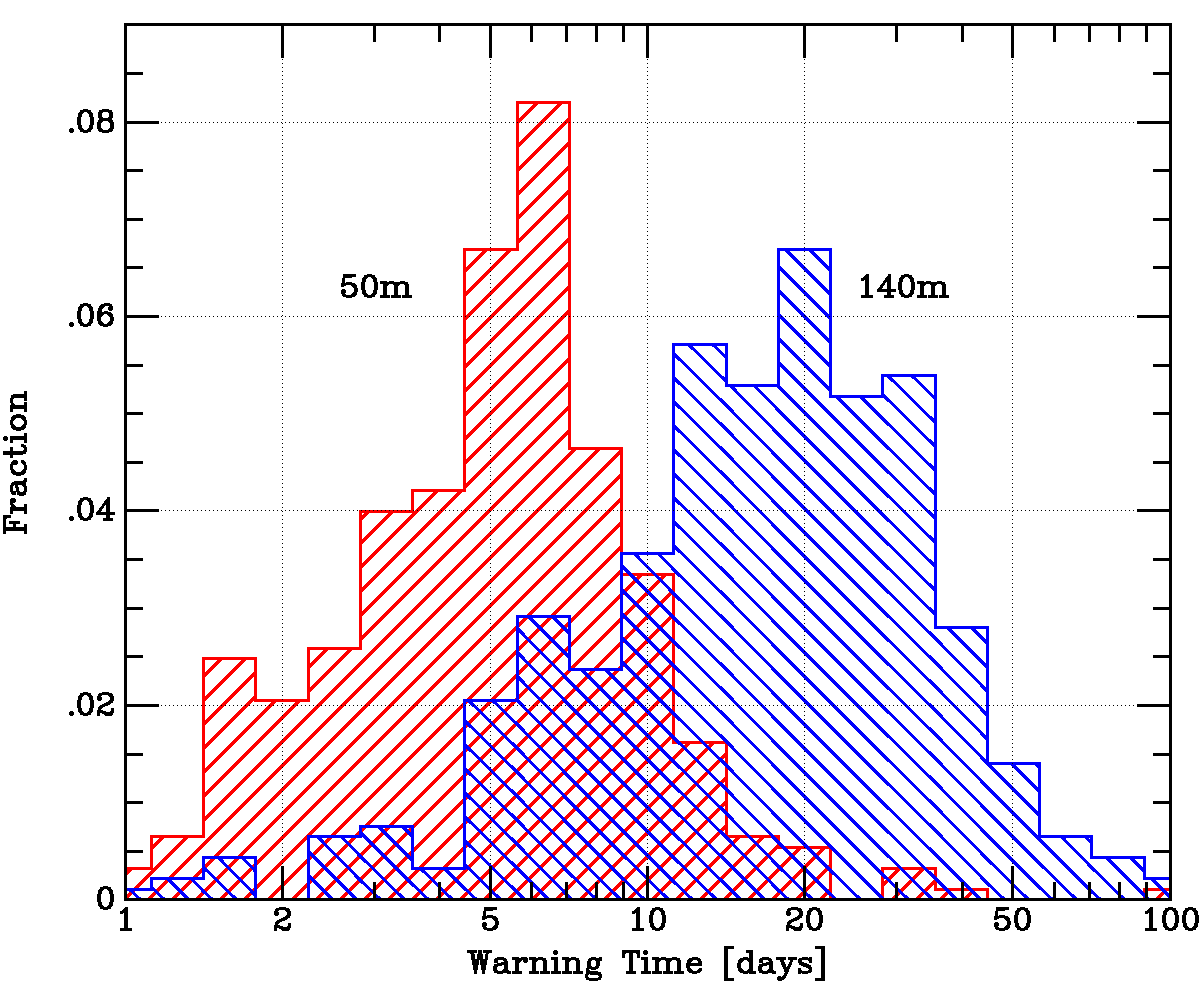}
            \includegraphics[scale=0.18]{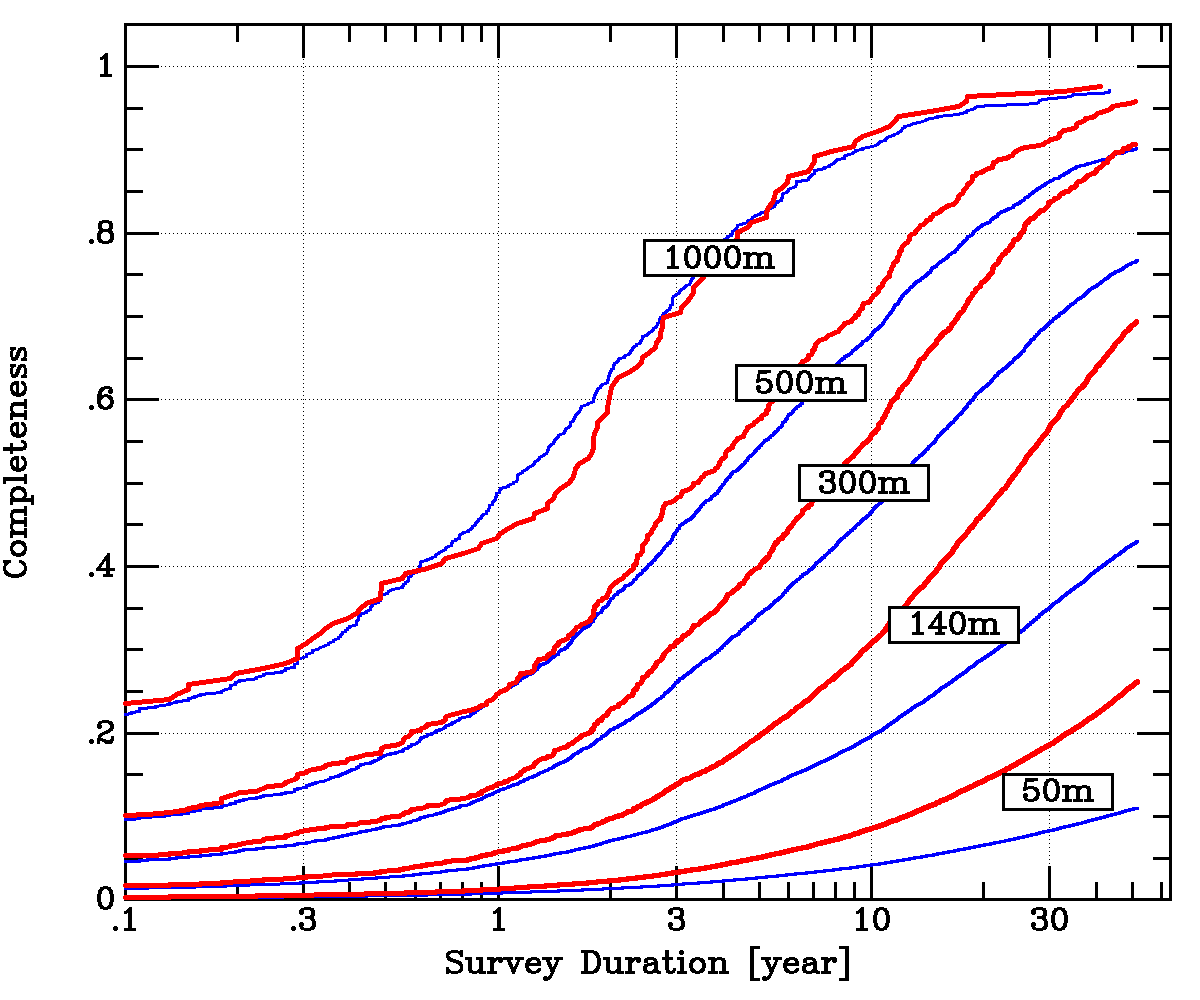}}
\caption{The left plot shows the distribution of warning times
  provided by ATLAS for impactors of 140~m and 50~m diameter that it
  detects.  The typical 140~m impactor will be
  found three weeks before arrival; the typical 50~m impactor will be
  found one week before collision.  Depending on survey duration,
  there is also a growing number of warning times longer than a year
  not illustrated here.  The right plot shows ATLAS's
  completeness for detection of NEOs (blue, thin lines) and PHOs (red,
  thick lines) of various sizes as a function of survey duration.
}
\label{fig:warning}
\end{center}\end{figure}

We have developed a detailed ATLAS simulator that integrates the
orbits of NEOs or
impactors from Veres et al (2009), either 10,000 impactors chosen to
strike the Earth randomly in location and time over the next 100
years, or the full population of 270,000 NEOs.  It is important to
note that Earth impactors have a different orbit distribution from
NEOs or even potentially hazardous objects (PHO's) --- the impactor's
orbit distribution is shifted to smaller semi-major axis, eccentricity
and inclination.\footnote{An NEO is defined as an object with
  perihelion less than 1.3~AU and aphelion greater than 0.983~AU; a
  PHO is an object with $H<22$ 
  (diameter $\sim$140~m) whose orbit passes within 0.05~AU of the
  Earth's orbit; an ``impactor'' is an object that actually strikes
  the Earth within 100 years.} This is what makes ATLAS so effective
at identifying {\it impactors} as opposed to generic NEOs.  If an
asteroid can hit the Earth, its orbit must intersect the Earth's orbit
and ATLAS's small search volume and fast cadence is ideal for finding
them.

The simulator uses ATLAS's view of each night's sky, schedules
the observing time, examines each asteroid for visibility according to
its apparent magnitude and the observation's extinction, trailing
losses, and weather, and decides that an asteroid has been ``found'' when
it has been observed 8 times in 4 tracklets or else has an accurate
parallax.

ATLAS can detect more than half of impactors larger than 50~m and
almost two thirds of 140~m impactors, as illustrated in
Figure~\ref{fig:warnfrac}.  The asteroids that ATLAS misses slip in
from the direction of the Sun and south pole or during
periods of bad weather. (An ATLAS copy in the southern hemisphere
or in a different weather pattern would raise the detection fraction.)
Figure~\ref{fig:warning} illustrates the distribution of warning times
provided by ATLAS.  Objects of 140~m diameter are typically
detected 20~days before impact while 50~m diameter objects are
detected with a week's notice, when they are $\sim$15 times the
distance to the moon.

Simultaneous images from the two sites provide a $3\sigma$ parallax
when the object is closer than 0.1~AU, about two weeks before impact.
The parallax is crucial for identification of an approaching asteroid in
a single night and important for vetoing confusion with space junk.

Looking at the full NEO and PHO populations,
the right panel of Figure~\ref{fig:warning} shows the rate at which
asteroids are found by ATLAS.  ATLAS will (re)discover 50\% of all
1000~m NEOs and PHOs within 1 year, 70\% in 3 years, and 90\% in 10
years.  The net rate of detection of 140~m asteroids or larger should
be more than 400 NEOs and 100 PHOs per year.  ATLAS can find about 15\%
of all 140~m PHOs within 3 years and 30\% within 10 years, slightly
less than what Veres et al (2009) found for the Pan-STARRS-1 mission.

ATLAS is not a direct competitor for the much more capable
surveys such as Pan-STARRS-4 or LSST.  However, for the
crucial days and weeks that an impactor is on final approach, ATLAS is
far more effective than any existing NEO survey, Pan-STARRS, or LSST.
The coverage and cadence that ATLAS provides gives us a high
probability of seeing an incoming asteroid, and ATLAS's sensitivity
is enough to spot it while it is still reasonably distant.

When ATLAS detects a nearby object it will automatically provide the
measurements to the Minor Planet Center for posting on their web-based
NEO Confirmation Page.  At that time, in a manner similar to the rapid
followup of the first pre-impact identification of meteoroid 2008
TC$_3$ (Jenniskens et al. 2009), we expect other amateur and
professional observers to obtain additional observations.  The case of
2008 TC$_3$ demonstrates the amazing accuracy 
that can be achieved: JPL's predictions were accurate to $\sim$20~sec
and $\sim$100~km within hours of discovery, and the eventual
prediction was accurate to $\sim$1.5~sec and $\sim$1~km.

%%%%%%%%%%%%%%%%%%%%%%%%%%%%%%%%%%%%%%%%%%%%%%%%%%%%%%%%%%%%%%%%

\subsection{Asteroid Science}

ATLAS will monitor a large number of asteroids in the Main Belt as
well as asteroids and comets elsewhere
in the Solar System.  The rapid time cadence of ATLAS is particularly
well suited to providing light curves and simultaneous colors for many
asteroids that can then be analyzed to infer asteroid shape and
tumbling motion.  ATLAS should achieve photometry in two colors with
0.01 mag accuracy at $V\sim16$, 0.02 mag accuracy at $V\sim17$, and
0.04 mag accuracy at $V\sim18$.  The AstDys web site
(hamilton.dm.unipi.it/astdys) lists 973 numbered asteroids with
$V<16$, 3,149 with $V<17$, and 10,078 with $V<18$ at this moment in
time (2010-10-10) in the 3/8 sky between RA 18$^h$ and 6$^h$ and Dec
$-30^\circ$ to $+90^\circ$.  Most are considerably off of opposition
right now, so have been or will be brighter during the 4--5 months it
takes to sweep by.  Therefore ATLAS should provide twice-nightly two
color light curves of $\sim$4 month duration for at least 2,000
asteroids at an accuracy of 0.01 mag per point, 6,000 asteroids with
an accuracy of 0.02 mag, and 20,000 asteroids with an accuracy of 0.04
mag.

It has been estimated by R. Jedicke (private communication) that there
is a collision each day that disrupts a 10~m asteroid in the Main
Belt.  If the dust cloud from the collision grows to 1000~m before
becoming optically thin the collision should be detectable by ATLAS.
The recent asteroid collision event P/2010 A2 discovered by LINEAR
achieved $m\sim19$ and would be detectable by ATLAS.

An object in the outer solar system must have $H{<4}$ to be detectable
by ATLAS, so ATLAS is unlikely to add many new discoveries to the
bodies already known or that will be discovered by Pan-STARRS1.  The
value from ATLAS is completeness and ongoing monitoring of 3/4 of the
sky.  If an object has slipped between the cracks of other surveys, or
happens to brighten (e.g. tumbling), or is rapidly approaching
(e.g. new comets) ATLAS may be the first to discover it.  The current
IAU definition of a dwarf planet is a body with $H{<}1$.  With full
illumination, at 60~AU distance (approximately the outer edge of the
Kuiper belt), such a body would have a $V$ magnitude of 18.8, and
therefore be easily detectable by ATLAS.  ATLAS should therefore
detect virtually all dwarf planets in the solar system within one
year, and be particularly useful for searching well out of the
ecliptic, where such bodies might have scattered.

\subsection{Supernovae}

SNIa have proven utility as measures of the cosmological expansion of
the universe, and it is hoped to continue with even more subtle
questions such as whether the accelerated expansion is consistent with
a cosmological constant.  Since we do not understand very well the
environment, initial conditions, trigger mechanism, and explosion
process of SNIa, these extremely delicate measurements are
vulnerable to systematic errors.  

A Type Ia supernova at $z=0.1$ peaks at $V=18.9$.  The
ATLAS sensitivity at SNR=10 is 19.5 per day; assuming 70\% clear
weather, the ATLAS sensitivity for 4 nights is $V=20.1$ for SNR=10.
According to Mannucci et al. (2007) there are some 9,000 SNIa yearly
closer than $z=0.1$ (32,000 at $z<0.15$).  Since the area that
ATLAS surveys each night is half of the entire sky (neglecting
obscuration by the galactic plane), we can expect that
ATLAS will find and follow 4,500 SNIa per year at $z<0.1$ and
SNR~$>30$ and 16,000 SNIa per year at $z<0.15$ with SNR~$>14$, with a
4 day sampling of the light curve.  Perhaps more interesting from the
standpoint of investigating systematics, ATLAS should find some 300
SNIa per year that peak at $V<17$, and $\sim$30 per year peaking at
$V<15$.  This is nearly an order of magnitude greater than the
discovery rate of bright SNIa over the past decade, and has the
advantage of being completely unbiased.  By contrast,
the KAIT telescope (Li et al, 2003) finds approximately 75 supernovae
per year by patrolling nearby galaxies.

The huge number of SNIa discovered
by ATLAS as well as the completeness of examining the entire sky will
empower us to ask questions such as characterization of explosion as a
function of host galaxy properties, details of the very early phases of the
explosion, and identification of outlier events that can be flagged
for spectroscopy or more detailed photometry.  We can also expect to
see a large number of SNIa in interesting environments such as rich
clusters or tidal streams of interacting galaxies.

Core collapse supernovae (CCSN) are particularly interesting when they
result in a huge, collimated explosion creating a gamma ray burst.
These are thought to occur in low metallicity environments from WR
stars that have a high core angular momentum at time of collapse.
Young et al. (2008) estimate that there are 20 CCSN per
year in galaxies with $12+\log(O/H)<8.2$ and $z<0.04$.
% in the 0.14 sky of the SDSS DR5.
ATLAS's nightly $10\sigma$ sensitivity of
$V=19.5$ translates to $V=18.5$ at $25\sigma$.  In 10 day's time
(7 clear nights) ATLAS achieves $V=19.6$ at $25\sigma$.  CCSN peak
at $M\sim-16.8$ (IIP), $M\sim-18.3$ (Ib/c), and $M\sim-19.6$ (IIL),
and the distance modulus at $z=0.04$ is $(m{-}M)=36.2$, so ATLAS
should be able to detect all of these outbursts at $25\sigma$,
extinction permitting.  Since ATLAS is surveying half the sky, the
expected number is 10 CCSN per year in such low metallicity galaxies.

\subsection{Gravity Waves}

Although LIGO has yet to detect a gravity wave (GW) event, it is
virtually certain that gravity waves exist and highly likely that
Advanced LIGO will detect GW events.  The most common detections will
be coalescing, compact objects whose changing quadrupole
moment makes a vigorous, detectable ``chirp'' in gravity waves.
Abadie et al. (2010) estimate the rates of events and sensitivities
and suggest that the most likely events will be neutron star - neutron
star coalescence within $\sim$445~Mpc.

While in principle a coalescence of naked compact objects could give
rise to minimal E\&M signature, it seems quite possible that the
release of more than $10^{53}$ ergs of energy might be accompanied by
$10^{48}$ ergs in the optical, as argued by Stubbs (2007).  Such an
explosion corresponds to a luminosity with absolute magnitude of
$M\sim-18$ for a duration of two days.  We do not try to advocate any
particular mechanism, but only make the point that if even a part in
$10^{-5}$ of the energy release appears in the optical, it will be a
substantial luminosity for a substantial duration.

Abadie et al. estimate that one NS-NS coalescence occurs every Myr
in every Mpc$^3$, and therefore expect to see some 40 events per year,
applying a factor of 2.26 to the horizon distance of 445~Mpc to
account for sky location and orientation.  The distance modulus of
445~Mpc is 38.2, so an explosion of $M=-18$ would be just detectable
by ATLAS, provided that it happens within the half sky visible to
ATLAS and endures long enough for ATLAS to sweep across it.  The
nearest object that Advanced LIGO would see in one year would be a
factor of $40^{1/3}$ closer, 2.7~mag brighter, and easily seen by
ATLAS.

The coalescence of compact objects creates a well defined GW signal
and an extremely ill-defined optical transient.  On the other hand,
core-collapse supernovae (CCSN) are common, and Ott (2009) has
reviewed their possible GW signatures and the ability of Advanced
LIGO to detect them.  Advanced LIGO may be able to detect a CCSN
within 1~Mpc with signal to noise of $\sim6$, but the rate of such
supernovae within the Local Group is only one in $\sim20$ years.
There is about one CCSN each year within $\sim5$~Mpc, and the rate
grows rapidly at distances beyond $\sim8$~Mpc that start to reach into
the Virgo cluster.

CCSN closer than the Virgo cluster will be much brighter than the
ATLAS magnitude limit of 20, and therefore ATLAS will detect the
half of them that are in the visible sky at time of explosion.  We
have a decent chance over a year or two of matching an Advanced LIGO
event at 3$\sigma$ with a CCSN seen by ATLAS, but there is no question
that ATLAS will provide times and locations for many events for which
Advanced LIGO may have a $\sim$1$\sigma$ detection.  The time between
collapse and emergence of the light flash is short enough that it
should be possible to correlate CCSN events with low SNR GW events and
thereby learn about the mechanism by which CCSN events create gravity
waves.

\subsection{Novae, Outbursts, and Variable Stars}

Novae range in absolute magnitude from $M_V = -9$ to $M_V = -7$,
declining by 3 mag in a week to several months, and the number per
galaxy is estimated to be some 40 per year.
ATLAS's 4 day $10\sigma$ sensitivity of $V=20.3$ gives
us the ability to see novae to distance moduli of $(m{-}M)\sim28$,
i.e. we will certainly see all the novae in the Milky Way and M31
within ATLAS's half sky, and most of the novae in nearby galaxies such
as M81 and M101, but we will not see novae in the galaxies in the
Virgo cluster.

Luminous blue variable stars flare at $M_V\sim-9$ to $M_V\sim-13$, so
again, ATLAS can find and monitor many of them on a daily basis to
distances as great as the Virgo cluster.

Mira variables peak at $M_V\sim-1.5$ with periods of order a hundred
days, and FU Orionis outbursts peak in the range of $M_V\sim-1$,
rising over a year and then declining over decades.  ATLAS can
monitor these all throughout the Galaxy, although M31 is too distant.
A southern ATLAS would encompass the Magellanic Clouds, at distance
modulus 18, so is sensitive at $10\sigma$ to $M_V=+2$ and brighter.

There are some 2000 cataclysmic variables known in the neighborhood of
the Sun.  ATLAS is unique in being able to keep an eye on all of them
with two samples each day spread by an hour or two.  This will provide
excellent sampling of their periods (typically about an hour), as well
as providing an alert within a day of an interesting outburst.  Within
1~kpc ATLAS has a $10\sigma$ sensitivity at $M_V=+9$ per visit.

Of course ATLAS will also watch the lesser beasts of the variable zoo
in the sky.  At 20~kpc ATLAS's sensitivity at $10\sigma$ per day is
$M=+3$, so all instability strip stars such as RR Lyrae and Cepheids
will have daily observations at high SNR.  ATLAS will provide the
first opportunity to catalog all the eclipsing and variable binary
stars in the sky to $M=+3$ or fainter.  ATLAS's blue filter is
deliberately truncated short of $H\alpha$, so ATLAS has special
sensitivity to $H\alpha$ flares --- variability that is extremely
``red'' is likely to arise from $H\alpha$.

\subsection{Active Galactic Nuclei}

Croom et al (2004) analyzed the AGN luminosity function
in the SDSS DR5, from which we calculate that there are some half million
AGN in the sky brighter than $V=19.6$.  At that level ATLAS can
monitor half of them, more than 100,000, at $25\sigma$ for a 10 day
cadence or $10\sigma$ for a daily cadence.  AGN have a complex
structure function of variability ranging from general flickering of
the accretion process to flares from tidal events to the spectacular
luminosities of blazars and their instabilities.  ATLAS's
sensitivity, solid angle, and cadence has an excellent overlap with
the densities of AGN and their various sources of variation.
Well sampled light curves with time scales ranging from days to years
are key to studying AGN variability.

At $V=20$ there are 1800 galaxies per square degree, so ATLAS will
maintain a daily watch on some 40 million galaxies, sensitive to events
whose luminosity is comparable to that of the galaxy on a 1 day
timescale or longer.  For example, 
a star is occasionally disrupted by accretion onto a black hole at the
center of a galaxy, producing a flare of predictable color and
duration.  Gezari et al. (2009) predict a volume rate
of $2.3\times 10^{-6}$~yr$^{-1}$Mpc$^{-3}$, and calculate that a survey
with $g<19$ would detect events out to 200 Mpc, corresponding to
detection rate of 20 events per year by ATLAS.

\subsection{Lensing}

Even at the north galactic pole there are about 2000 stars per square
degree brighter than the 1 day limiting magnitude of $V=20$.  Over
the entire 20,000 sq deg being surveyed each night by ATLAS there are
about a half billion stars far enough off of the galactic plane that
there is manageable confusion in ATLAS's 4.4$^{\prime\prime}$ pixels.

Han (2008) estimates the rate of near-field microlensing from all sky
surveys and finds that at $V=18$ we can expect to see 23 events per
year, where an event is defined as an increase of source flux by more
than 0.32 mag.  The number scales as $10^{0.4m}$, so at $V=19$ there
should be 58 lensing events over the sky per year.  At $V=18$ the
lens-source proper motion can exceed 40~mas~yr$^{-1}$ so we could hope
to disentangle their light after a few years by imaging from space or
ground-based adaptive optics.  The event timescales at the fainter
limits are about 20 days.

ATLAS will survey half the entire sky at SNR 10 per night at
$V=19.5$, so a $V=18$ star will be captured with 0.025 mag uncertainty
each night and a $V=19$ star will have 0.06 mag error each night.  A 0.3
mag lensing event will therefore be seen at $10\sigma$ at $V=18$ and
$5\sigma$ at $V=19$.  Over 20 nights, even allowing for weather,
ATLAS ought to capture most events to $V<19$, and the two filters
will permit some level of testing of achromaticity.  We therefore
estimate that ATLAS should see approximately 30 microlensing events
in the near-field each year, and 10 should have high signal to noise
and time coverage.

There is a suprisingly large cross section for strong gravitational
lensing by galaxy centers.  We integrated the galaxy velocity
dispersion function of Sheth et al. (2003), using both
the densities listed for early-type galaxies and late-type galaxy bulges,
assuming that each galaxy has an isothermal core capable of lensing.
Over the entire sky, the cross section for a lensing magnification of
3 or greater is 0.37~deg$^2$ for sources at a redshift of $z=0.5$,
1.75~deg$^2$ for sources at a redshift of $z=1.0$, etc, scaling as
$(\mu-1)^{-2}$, where $\mu$ is the lensing magnification, and as $z^3$
in the Euclidean limit.  Divided by the 41,250~deg$^2$ of the sky, this
cross section provides a magnification probability.

The number of lensed SNIa that ATLAS will see, even aided by
magnification, will barely yield one event per year.  Integrating the
density of SNIa from Mannucci et al (2007) against the lensing cross
section, we expect to see one lensing event per year peaking at a
(magnified) magnitude of 20.6.  Of course this will be extremely hard
to distinguish from the hordes of SNIa close to galaxy centers.

However, the number of magnified AGN that ATLAS will detect is quite
large.  Integrating the luminosity function of Croom et al. against
this cross section gives more than 40 AGN in the sky with a
magnification of 3 or larger at a magnitude of 19.6 or brighter
($25\sigma$ at 10 days, $10\sigma$ at 1 day).  Among these there are
$\sim7$ at magnification 10 or greater.  The lensing cross section
grows rapidly with redshift, of course, so this is an interesting
probe of AGN density, AGN variability as a function of redshift,
and galaxy core structure as a function of redshift.  At high
magnifications the microlensing by stars within the lensing galaxy can
cause substantial flickering as well.

\subsection{The Static Sky}

\begin{figure}[!hb]\begin{center}
\centerline{\includegraphics[scale=0.25]{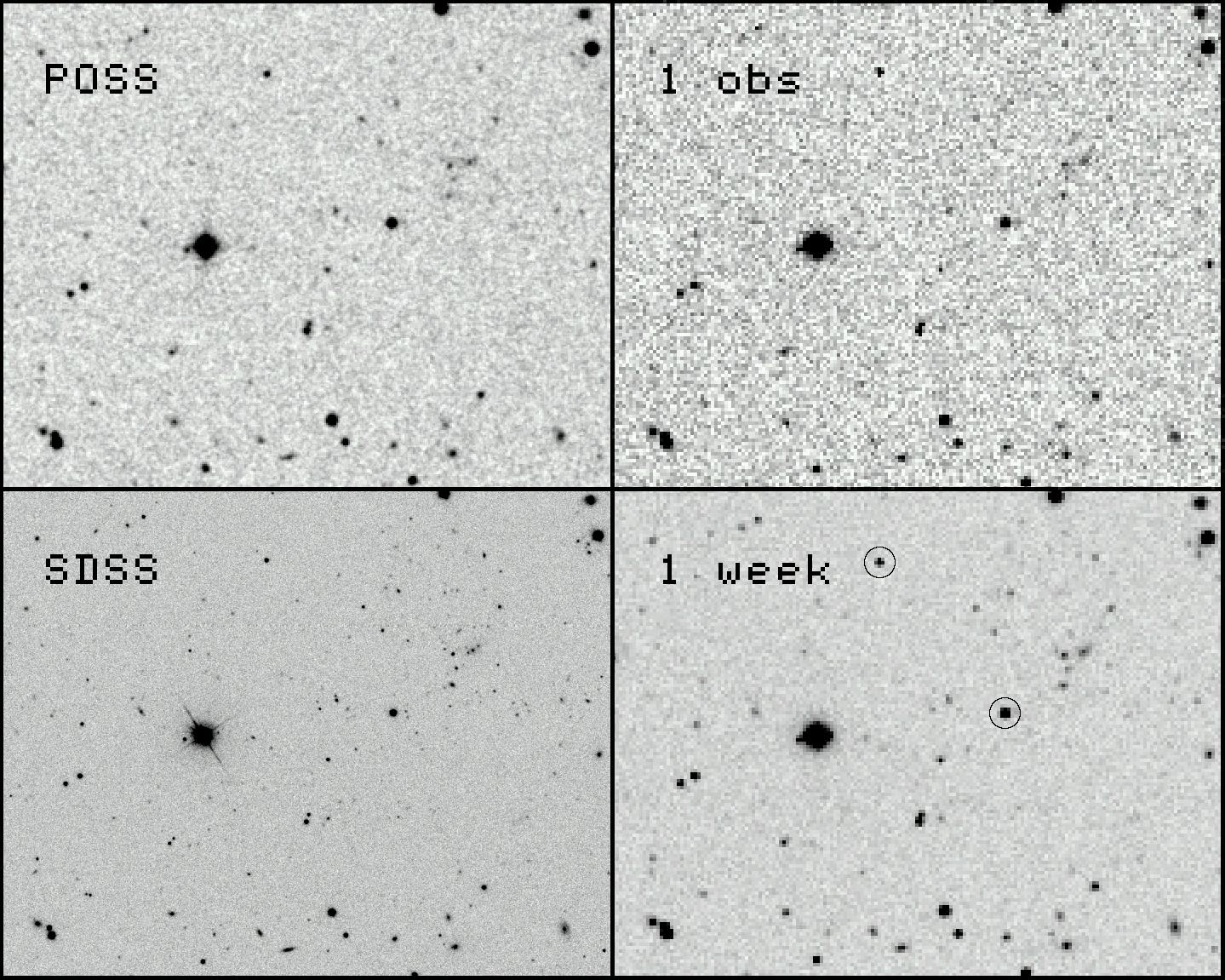}}
\caption{The appearance of an $0.1\times0.08$ deg portion of the sky
  in POSS $R$ band is shown in the upper left, the view in the SDSS $r$
  band in the lower left, how it would appear in a single ATLAS red
  observation (two 30 sec integrations) in the upper right, and after
  a week of observation (5 clear nights) in the lower right.  The
  circled stars are $m=17$, $m=18.5$, $m=19.4$ (5$\sigma$), and a
  pair at $m=18.2$ and 19.2 separated by 4.4\arcsec.}
\label{fig:poss}
\end{center}\end{figure}

\begin{figure}[!hb]\begin{center}
\centerline{\includegraphics[scale=0.5]{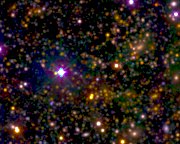}}
\caption{A color image of the $0.1\times0.08$ deg portion of the sky
  after a year of ATLAS observation illustrates the appearance of
  SNR $10\sigma$ at $V=22.4$.  Although static objects are starting
  to blend together at this limit, variable objects are rare enough
  to stand out cleanly once the static sky is subtracted.}
\label{fig:dpsisky}
\end{center}\end{figure}

ATLAS has only modest sensitivity on a per-unit telescope basis, and it
undersamples a point spread function that will average about
2.5\arcsec\ from atmosphere and optics, but the SNR grows as
each piece of the sky gets some 350 visits per year (3.1 mag).
Figure~\ref{fig:poss} compares how an $0.01$~sq~deg
piece of sky appears in the digital POSS sky survey, the SDSS sky
survey, and ATLAS.
ATLAS goes substantially deeper than POSS after a week of
observation, even allowing for weather.  The SDSS PSF is
considerably better than ATLAS can ever achieve, but
Figure~\ref{fig:dpsisky} illustrates how the color co-added sky would
look after a year of ATLAS observation (with allowances for weather).
After a year's operation ATLAS is significantly deeper than SDSS (and
covers 3/4 sky), although the coarse sampling eventually runs into source
confusion.  However, the density of transient and variable objects is
far lower than the density of static objects, so ATLAS should stay
above the confusion limit for detection and characterization of the
variable parts of their light curves, with sensitivity to timescale
$\tau$ that goes as $\tau^{1/2}$.  For example, ATLAS can achieve
$m\sim23$ sensitivity for slow events such as AGN variability or
long-period variable stars.

\subsection{Space Junk}

There is a growing amount of ``space junk'' in low Earth orbit (LEO)
and geosynchronous Earth orbit (GEO).  This is of some concern for
satellites and space travel, as evidenced by the recent destruction of
an Iridium satellite by collision with a tumbling booster.  To sky
surveys looking for transients and moving objects, space junk is a
significant background signal that masks the solar system objects we
are interested in.

A good rule of thumb for the detection of a streak across an image
left by a moving object is that when each PSF-sized segment of the
streak is $1\sigma$ above background it is easily visible to the eye
and easily detectable.  A magnitude fainter is harder, and two
magnitudes fainter is just about the limit of what can be detected.

A LEO object at 1000~km range moves at about 0.4~deg/sec.  ATLAS has
four views, two staggered by 5~deg at each site and two sites, so many
objects will have one or two endpoints caught by the system.  A LEO
object will spend 3~msec on a pixel, and therefore the $1\sigma$
routine detection benchmark is 8.3 mag brighter than the $5\sigma$
magnitude limit for a 30 sec exposure, i.e. about $m=10.8$.  At
$m=11.8$ the streak will be visible to the eye and should be possible
to detect automatically, particularly given the confirmation from the
two sites and the possible collection by the adjacent pointing.

This magnitude corresponds to a white, fully illuminated Lambertian
ball of size $\sim$4~cm or a piece of space junk of size $\sim$15~cm
of albedo 0.1 and random illumination.  There are estimated to be of
order 10,000 objects in LEO of that size or greater.

A GEO object only moves at 15\arcsec/sec, spending 0.3~sec on a pixel,
so the 1-sigma benchmark is 5 mag fainter than LEO,
i.e. $m=15.8$.  Automated detection should be possible therefore to
$m=16.8$ ($\sim$60~cm) without much trouble.  At 40,000~km range, the
tangential position accuracy should be approximately 100~m, and the
range accuracy about 50~km for the single observation.  Since ATLAS 
sweeps the visible sky twice each night, GEO objects will be
captured on two occasions, permitting a good estimate of orbit.

\section{World-wide Internet Survey Telescope}

Apart from its value as a survey for hazardous asteroids, ATLAS also
serves to define a unit survey telescope and software that can be
replicated many times in order to improve on the networked
performance.  ATLAS could therefore be a template for a ``World-wide
Internet Survey Telescope'' (WIST) that has as many unit observatories
as there are parties who would like to participate, since it is sized
to fit within the budget of any college or university.

The ATLAS proposal to NASA is carefully optimized to
take advantage of the existing $\sim$\$1.3~M of detectors, tries to
avoid as much hardware risk as possible by using commercial
components, and plans for software R\&D to be the most costly,
difficult, and critical aspect of the project.  However, given time and
resources for hardware R\&D, it is possible to develop a successor
that would have considerably greater \merit\ per unit cost than the
present ATLAS implementation.  

For example a remarkable set of designs by M. Akermann, J.T. McGraw,
and P.C. Zimmer (private communication) include an 0.5~m aperture
Hamilton astrograph with 1~m focal length and 7~deg FOV that achieves
50\% encircled energy at 1.5~um radius, i.e. half-energy within a
diameter of 0.6\arcsec.  The optimal unit camera could have 5--9~um
pixels (0.5--0.2~Gpixel total) for a 1--1.8\arcsec\ sampling over a
field of view of $\sim$40~sq~deg.  We note the growing availability
of large format detectors, such as the STA-1600 CCD
(www.sta-inc.net) that has $10\times10$k 9~um pixels and can read out at 1
frame per second, Canon's recent announcement of an APS-H
format (29$\times20$~mm), 120~Mpixel CMOS sensor as well as a
monolithic 202$\times$205~mm, high-sensitivity sensor, and advances in
% www.dpreview.com/news/1008/10083101canonlargestsensor.asp
back-illuminated CMOS detectors by Cypress, Sony, and others.  It is
not inconceivable that improvements in \merit\ per unit cost of a
factor of three or more might emerge.

We have therefore tried to think of ATLAS as the start of a
``franchise'' that defines what a hardware and software unit should
be, basically a high performance telescope and detector system in a
robotic observatory, with common reduction and analysis pipeline and
common protocols for communications.  An essential component is
bandwidth management -- the ATLAS system has enough local processing
to handle the $\sim$100GB per night of raw data, and depends on
extra-observatory bandwidth primarily for trading object catalogs that
will be orders of magnitude less information.

The Las Cumbres Observatory Global Telescope (Brown et al. 2006)
consists of a widely distributed set of many telescopes that are
intended for full-time synoptic coverage of interesting events such as
planet occultations or microlensing passages.  WIST differs by being
dedicated to all-sky, nightly survey and discovery.  WIST and LCOGT
are therefore complementary in their approaches and scientific goals,
although both are striving to make the greatest possible use of the
silicon revolution and the internet.  

Because we want to achieve the highest possible hardware and software
performance, and also because we cannot cope with the difficulties
inherent in a ``National Virtual Observatory'' that tries to accept
information of varying quality and provenance, the ``WIST franchise''
would insist on a very high degree of standardization.  This does not
mean that a college with poor seeing or bright skies cannot
participate.  WIST is intended to be tolerant of poor seeing, and the
standardization allows us to emplace metrics that permit optimal
combination of information from all sites.

As the number of observatories grows it would become possible to
schedule dynamically according to weather, and to allocate
observations by filter or time to different places.  Sites with very
bright sky backgrounds might be assigned very short exposure times for
a more rapid cadence on bright objects, for example.  Obviously the
search for approaching impactors can be significantly improved by
WIST, by virtue of closing the southern blind spot, squeezing down the
solar blind spot, immunity to weather, and by deeper imagery.

Since we intend that the cost of WIST be borne by the participants, we
expect that all results would be made public immediately.  This will
encourage participants to define and carry out projects promptly, and
we hope will encourage collaborations.  The fact that WIST puts the
smallest colleges on the same footing with the most prestigious
research universities will serve to foster innovation and reduce the
dependence of research achievement on availability of resources.

\section{Conclusion}

In this article we have argued that the congressional mandate to NASA
to mitigate the hazard from asteroid impact on Earth can be parsed
temporally as well as by impactor size.  To some degree the risk
can never be reduced to zero because orbits are continually perturbed,
but we believe that the most important {\it time interval} for
discovery of small impactors is just before impact, and we have
demonstrated that it is relatively easy and cost-effective to patrol
that portion of hazard space.

In order to understand the real capability of survey systems we
examined the meaning of ``etendue'' and survey \merit, and derived
equations that quantify survey capability both in terms of hardware
specifications of aperture, solid angle, efficiency, and image
quality, but also in terms of actual performance.  In support of this
we also derived an equation that quantifies the image performance in
terms of an ``effective noise footprint'', valid even in the regime of
pixels that undersample the PSF.

We advanced a description of a new survey instrument, called the
``Asteroid Terrestrial-impact Last Alert System''.
ATLAS surveys 20,000 sq deg twice per night to magnitude 20, the
sensitivity determined by the practical need for three week's warning
of a 100~Mton impact, and the solid angle determined by the practical
need for immediate warning of a 1~Mton impact.  The components for an
ATLAS system are mostly available commercially, the cost is low, and the
construction time short.

We compared ATLAS with the other prominent sky-surveys in operation or
being planned, and found that ATLAS has interesting complementarities
to the other surveys that make it particularly potent for early
warning of impacts by hazardous objects: it surveys essentially the
entire sky at a rapid cadence so the probability of any object
slipping through is reasonably low and its sensitivity is high enough
to provide a useful warning.  No other survey is as effective for this
particular job.

Of course the ATLAS imagery will also open an interesting window on
the entire transient universe.  Some of the additional science
products that ATLAS will produce include:
\begin{packed_item}
\item hundreds of light curve points for thousands of asteroids that
  provide estimates of shape and spin,
\smallskip
\item detection of all dwarf planets in the solar system,
\smallskip
\item twice per night monitoring of $\sim$2000 cataclysmic variables,
\smallskip
\item detection of $\sim$30 near-field microlensing events per year,
\smallskip
\item twice per night light curves of most variable stars in the Galaxy, 
\smallskip
\item novae and luminous blue variable outbursts in most nearby galaxies,
\smallskip
\item detailed and prompt information on E\&M counterparts to
  gravity wave events,
\smallskip
\item detection of $\sim$10 core collapse SN in 
low metallicity galaxies per year,
\smallskip
\item light curves of $\sim$4000 SNIa per year with $z<0.1$, $\sim$300
  with $V<17$, and
\smallskip
\item variability of $\sim$100,000 AGN.
\end{packed_item}

We ended with consideration of a future ``World-wide Internet Survey
Telescope'' (WIST), comprised of a confederation of many basic
observatory units.  The intent is to ``franchise'' the hardware and
software design, but at a cost point that makes an observatory unit
affordable by essentially any college or university.  This
standardization is necessary to allow the full system to be scheduled
and for the results of the imagery to be optimally combined, but leads
to a system that is virtually unbounded in its ability to explore the
time and sensitivity domain of the transient universe.

We acknowledge useful discussions with Christopher Stubbs, Robert
Jedicke, Armin Rest, and John Blakeslee.  We are grateful for the
remarkable design work of Mark Ackermann, John McGraw, and Peter
Zimmer.  This paper benefited from discussions at the Aspen Center for
Physics.  Support for this work was provided by National Science
Foundation grant AST-1009749.


\begin{thebibliography}{Author(1982)}

\bibitem[Abadie, J. et al.(2010)]{arXiv:1003.2480v2} Abadie, J. et
  al.\ 2010, Classical and Quantum Gravity, 27, 17, 173001

\bibitem[Asphaug, E. (2009)]{2009AnnRevEarthPlanSci..37..413} Asphaug, E.
\ 2009, Annual Reviews Earth and Planetary Science, 37, 413

% Predictions for the Rates of Compact Binary Coalescences Observable
% by Ground-based Gravitational-wave Detectors

\bibitem[Bakos et al.(2009)]{2009IAUS..253..354B} Bakos, G., et al.\ 2009, 
IAU Symposium, 253, 354 

\bibitem[Boslough \& Crawford (2008)]{2008JImpEng..35..1441} Boslough,
  M.B.E. \& Crawford, D.A.\ 2008, J. Impact Engineering, 35, 1441

\bibitem[Bottke et al.(2002)]{2002Icar..156..399B} Bottke, W.~F., 
Morbidelli, A., Jedicke, R., Petit, J.-M., Levison, H.~F., Michel, P., 
\& Metcalfe, T.~S.\ 2002, Icarus, 156, 399 

\bibitem[Brown et al.(2002)]{2002Natur.420..795B} Brown, M.~E., Bouchez, 
A.~H., \& Griffith, C.~A.\ 2002, \nat, 420, 795 

% \bibitem[Brown et al.(2006)]{2006AAS...208.5605B} Brown, T.~M., et
%   al., 2006, Bulletin of the AAS, Vol. 38, 136

\bibitem[Burgett \& Kaiser(2009)]{2009amos.confE..39B} Burgett, W., \& Kaiser,
  N.\ 2009, Proceedings of the Advanced Maui Optical and Space
  Surveillance Technologies Conference, held in Wailea, Maui, Hawaii,
  September 1-4, 2009, Ed.: S.~Ryan, The Maui Economic Development
  Board., p.E39

\bibitem[Croom et al.(2004)]{2004MNRAS.349.1397C} Croom, S.~M., Smith, 
R.~J., Boyle, B.~J., Shanks, T., Miller, L., Outram, P.~J., 
\& Loaring, N.~S.\ 2004, \mnras, 349, 1397 

\bibitem[Gezari et al.(2009)]{2009ApJ...698.1367G} Gezari, S., et al.\ 
2009, \apj, 698, 1367 

\bibitem[Han(2008)]{2008ApJ...681..806H} Han, C.\ 2008, \apj, 681, 806 

\bibitem[Ivezic et al.(2008)]{2008arXiv0805.2366I} Ivezic, Z., Tyson, 
J.~A., Allsman, R., Andrew, J., Angel, R., 
\& for the LSST Collaboration 2008, arXiv:0805.2366 

% \bibitem[Jenniskens et al.(2009)]{2009Natur.458..485J} Jenniskens, P. 
% et al., 2009, Nature, 458, 485.

\bibitem[Keller et al.(2007)]{2007PASA...24....1K} Keller, S.~C., et al.\ 
2007, Publications of the Astronomical Society of Australia, 24, 1 

\bibitem[Larson et al.(2003)]{2003DPS....35.3604L} Larson, S., Beshore, E., 
Hill, R., Christensen, E., McLean, D., Kolar, S., McNaught, R., 
\& Garradd, G.\ 2003, Bulletin of the American Astronomical Society, 35, 982 

\bibitem[Larson et al.(2007)]{2007AJ...133.1247L} 
Larsen, J., Roe, E., Albert, E., Descour, A.,
McMillan, R., Gleason, A., Jedicke, R., Block, M.,
Bressi, T., Cochran, K., Gehrels, T., Montani, J.,
Perry, M., Read, M., Scotti, J., and Tubbiolo A.\ 2007, \aj, 133, 1247

\bibitem[Law et al.(2009)]{2009PASP..121.1395L} Law, N.~M., et al.\ 2009, 
\pasp, 121, 1395 

\bibitem[Li et al.(2003)]{2003PASP..115..844L} Li, W., Filippenko, A.~V., 
Chornock, R., \& Jha, S.\ 2003, \pasp, 115, 844 

\bibitem[McMillan et al.(2006)]{2007prociau.S236} 
McMillan, R. S., and the Spacewatch Team\ 2006, 
Proceedings of the International Astronomical Union, Volume 2,
Symposium S236, August 2006, ``Near Earth Objects, our Celestial
Neighbors, Opportunity and Risk'', Published online by Cambridge University
Press 03 May 2007. Edited by: Andrea Milani, University degli
Studi, Pisa; Giovanni B. Valsecchi, University degli Studi Roma
Tre; David Vokrouhlicky, Charles University, Prague, pp 329-340

\bibitem[Malek et al.(2009)]{ProcSPIE..7502}
Malek, K., et al.\ 2009,
Proceedings of the SPIE, Vol. 7502

\bibitem[Mannucci et al.(2007)]{2007MNRAS.377.1229M} Mannucci, F., Della 
Valle, M., \& Panagia, N.\ 2007, \mnras, 377, 1229 

\bibitem[Melosh \& Collins(2005)]{2006NAT.434.157} Melosh, H.J. \&
  Collins, G.S. \ 2005, \nat, 434, 157

\bibitem[Morbidelli et al. (2002)]{2002Icarus.158.329} Morbidelli,
  A. et al. \ 2002, Icarus, 158, 329

\bibitem[NASA NEO Report (2007)]{2007NASA..NEO} 
NASA NEO Report 2007,
http://www.nasa.gov/pdf/171331main\_NEO\_report\_march07.pdf 

\bibitem[NRC Report (2010)]{2010NRC..NEO} 
NRC Report, 2010, ``Defending Planet Earth: Near-Earth Object Surveys
and Hazard Mitigation Strategies'', National Academies Press, ISBN 0-309-14969

\bibitem[Ott, C.D. (2009)]{2009CQGra..26f3001O} Ott, C.D.\ 2009,
  Classical and Quantum Gravity, 26, 6, 063001
% TOPICAL REVIEW:  The gravitational-wave signature of core-collapse
% supernovae
% archivePrefix = "arXiv:0809.0695

\bibitem[Pollacco et al.(2006)]{2006PASP..118.1407P} Pollacco, D.~L., et 
al.\ 2006, \pasp, 118, 1407 

\bibitem[Sheth et al.(2003)]{2003ApJ...594..225S} Sheth, R.~K., et al.\ 
2003, \apj, 594, 225 

\bibitem[Schechter et al. (1993)]{Schechter93} Schechter, P., Mateo,
  M., and Saha, A.,\ 1993, \pasp, 105, 1342
%title Dophot

\bibitem[Stokes et al.(2000)]{2000Icar..148...21S} Stokes, G.~H., Evans, 
J.~B., Viggh, H.~E.~M., Shelly, F.~C., 
\& Pearce, E.~C.\ 2000, Icarus, 148, 21 


\bibitem[Stubbs(2008)]{2008CQGra..25r4033S} Stubbs, C.~W.\ 2008, Classical 
and Quantum Gravity, 25, 184033 

\bibitem[Trujillo et al.(2001)]{2001MNRAS.328..977T} Trujillo, I., Aguerri, 
J.~A.~L., Cepa, J., \& Guti{\'e}rrez, C.~M.\ 2001, \mnras, 328, 977 

\bibitem[Vere{\v s} et al.(2009)]{2009Icar..203..472V} Vere{\v s}, P., 
Jedicke, R., Wainscoat, R., Granvik, M., Chesley, S., Abe, S., Denneau, L., 
\& Grav, T.\ 2009, Icarus, 203, 472 

\bibitem[Vestrand et al.(2003)]{2003AIPC..662..550V} Vestrand, W.~T., 
Albright, K., Casperson, D., Fenimore, E., Ho, C., Priedhorsky, W., White, 
R., \& Wren, J.\ 2003, Gamma-Ray Burst and Afterglow Astronomy 2001: A
Workshop Celebrating the First Year of the HETE Mission, 662, 550 

\bibitem[Young et al.(2008)]{2008A&A...489..359Y} Young, D.~R.,
  Smartt, S.~J., Mattila, S., Tanvir, N.~R., Bersier, D., Chambers,
  K.~C., Kaiser, N., \& Tonry, J.~L.\ 2008, \aap, 489, 359 

\end{thebibliography}
\end{document}